\documentclass[
 reprint,aps,
]{revtex4-2}

\usepackage{graphicx,amsmath}
\usepackage{epstopdf, epsfig}
\usepackage{mathrsfs,hyperref}

\usepackage{dcolumn}
\usepackage{bm}

\usepackage{amsmath,amssymb,color}
\usepackage{physics,mathtools}
\usepackage{siunitx}

\graphicspath{{Figures/}}

\newcommand*{\Oh}{\mathcal{O}}
\newcommand{\myeqref}[2]{(\hyperref[#1]{\ref*{#1}#2})}
\newcommand{\myref}[2]{\hyperref[#1]{\ref*{#1}(#2)}}
\newcommand{\myrefnb}[2]{\hyperref[#1]{\ref*{#1}#2}}
\newcommand{\subtag}[1]{\tag{\theparentequation #1}}

\newcommand*{\ee}{\hat{\bm{e}}}

\begin{document}

\title{Geometric dependence of curvature-induced rigidity}

\author{Hanzhang Mao$^1$}
\author{Thomas G.\ J.\ Chandler$^1$}
\author{Mark Han$^2$}
\author{Saverio E.\ Spagnolie$^{1}$}

\affiliation{$^1$ Department of Mathematics, University of Wisconsin-Madison, 480 Lincoln Dr., Madison, WI 53706}%
\affiliation{$^2$ Department of Physics, University of Wisconsin-Madison}

\date{\today}

\begin{abstract}
Bending the edge of a thin elastic material promotes rigidity far from its clamped boundary. However, this curvature-induced rigidity can be overwhelmed by gravity or other external loading, resulting in elastic buckling and large deformations. We consider the role of body geometry on this competition using experiments, numerical simulations, and reduced-order models. Finite element simulations are performed using a model nonlinear hyperelastic material, and a theoretical framework is proposed that incorporates small lateral curvatures, large longitudinal rotations, and a varying cross-sectional width. A particular focus is on the comparison between rectangular and triangular sheets, and trapezoidal sheets in between. Sheet geometry affects downward tip deflection by changing the relative importance of the sheet's weight and the rigidity provided by curvature, often in subtle ways. In extreme cases, non-monotonic deflection is observed with increasing sheet length, and a region of hysteretic bistability emerges, becoming more pronounced with  rectangular sheets and large imposed curvatures. These findings demonstrate the profound impact of geometry on the competition between curvature-induced rigidity and gravity-induced deformation in thin elastic materials.
\end{abstract}

\maketitle

A thin elastic sheet is granted a surprising degree of rigidity just by providing transverse curvature along one edge \cite{prkmct16,tbnv19}. This curvature-induced rigidity is commonly explained to be a consequence of Gauss's \textit{Theorema Egregium}, which states that the Gaussian curvature of a smooth surface is invariant under stretch-free deformations \cite{ap00}. Surface curvature imposed transversely on an otherwise flat, inextensible sheet thus restricts longitudinal bending, since the Gaussian curvature is non-zero at a saddle point. Loss of rigidity, which may occur in a catastrophic buckling event, thus requires material stretching. In practice, since stretching energy scales with the sheet thickness, $h$ and bending energy only with $h^3$, thin materials often remain nearly stretch-free and can benefit from the rigidity above.

Nature provides countless examples of passive and active stiffening using curvature (Fig.~\myref{Figure1}{a-c}). Familiar examples include leaves \cite{td16}, fish fins \cite{Bainbridge63,gjl94,aml07,nybvm17}, tongues \cite{gngw07}, the human foot \cite{vyedsthbm20}, and in the snapping mechanism of the Venus flytrap \cite{fsdm05}. The ability to dynamically control stiffness only by actions on the clamping boundary is also of industrial significance. Applications include the use of curvature-induced rigidity in flapping wings \cite{tdhmldbm07,mg18,mkl22}, flexible electronics \cite{rjfs21,bxyz23}, and in soft robotics \cite{lgfz21}, where stiffness response to external stimuli is a basic need \cite{mcc16}. In the latter application, stiffness is commonly controlled by material jamming \cite{nvh18,fdh20}, which is achieved by various means \cite{lzzpm19,wkn21,gmklwjasr23,byrk23}. The deployment and retraction of structures, on Earth and in space, also leans on strategic use of curvature \cite{mp20}, including the use of bistable tape springs \cite{sp99, clzhty22,tjas24}. Curvature-induced rigidity appears in an even more familiar setting as well --- folding a pizza at its crust is commonly done during eating to hold it aloft. Natural questions that arise in these settings include: What is the vertical deflection of the material under loading? How far does an imposed curvature persist into the material? And, more generally, how do equilibrium configurations and stability depend on  geometric and material properties? 

Previous authors have noted limitations on curvature-induced rigidity for elastic sheets of finite thickness \cite{btqf14,tbnv19}. A thin rectangular sheet which is curved and clamped along one edge, and subject to gravitational loading, tends to display one of two displacement modes. The first is a curvature-dominated mode, where the sheet is held aloft against gravity; the second is a gravity-dominated mode, where the boundary curvature is insufficient to prevent a rapid vertical drop. Which mode is selected depends on the length of the sheet, $L$, relative to a persistence length, $L_p$, the distance along the sheet over which the transverse curvature relaxes. In the absence of gravity, Barois et al.~\cite{btqf14} found the persistence length to be $L_p^0\sim a^2/\sqrt{h R}$, where the superscript indicates the lack of gravity, $a$ is the sheet width, and $R$ is the imposed radius of curvature at the clamped boundary. This $L_p^0$ was found to be accurate provided the sheet is sufficiently curved, $a^2/(8R h)\gtrsim 5$, and is notably independent of the Young's modulus of the sheet, $E$. 

Upon the introduction of gravity, Taffetani et al.~\cite{tbnv19} found the same scaling for short, curvature-dominated sheets, specifically when $L\ll E h^2/(\rho g R^2)$, with $\rho$ the material density and $g$ the gravitational acceleration. Upon increasing the sheet length, a buckling transition occurs at a critical `standout' length, $L=L_1^*$, due to the curvature-induced rigidity no longer supporting the sheet's weight. Far beyond this standout length, for $L\gg E h^2/(\rho g R^2)$, a separate scaling for the persistence length for long, gravity-dominated sheets was found of the form $L_p^g\sim[E/(\rho g L)]^{1/4}a^2/R$.

While it is natural to consider the buckling transition from curvature- to gravity-dominated deformations as the sheet's length is increased, decreasing the length also yields a critical `pop-up' length, $L=L_0^*$, at which this transition is reversed --- the sheet suddenly jumps back to its hoisted, curvature-dominated shape. As we shall see, the pop-up and standout lengths are not necessarily equal ($L_0^*\neq L_1^*$) and depend on the imposed curvature. An intermediate region thus emerges, $L_0^*\leq L\leq L_1^*$, where two equilibrium configurations may exist for the same sheet.

Equilibrium configurations and stability are commonly studied features of elastic materials, and have been observed and investigated in various related settings \cite{as15, ws18,holmes19,bkrw18}. Active control of buckling and stability can be achieved through a variety of mechanisms, including lateral swelling \cite{dhs11} and temperature response \cite{khbsh12}. In particular, snapping and buckling within the context of elastic surfaces \cite{hc07,vella19, dvkm07} have been discovered and investigated in cases involving intrinsic curvature \cite{psjh17, mbjr17, vm08}, induced curvature arising from external constraints or forces \cite{ap00,rpmkct18,bdcadvd23,lhrdytll23, scph19,hvhh24}, folded sheets \cite{da14,swg19,ws19,fdzbb24, wlfd24, yadh22}, and twisted sheets and ribbons \cite{cb08,ck13,sw19,khbkr20}. Our problem lies between these settings: while the curvature is induced by external forces, the gravity-induced buckling resembles that observed in structures with intrinsic curvature, such as tape springs \cite{tbnv19}.

These stability features are influenced by factors such as external constraints on boundaries \cite{wlxjdhzc20, bvv18, bvv16}, intrinsic curvature \cite{pssbh18}, and twisting \cite{gt37}. Perhaps surprisingly, curvature-induced rigidity can in some cases be enhanced by softening the clamping condition \cite{qb25}. However, a simpler, and perhaps more generic, geometrical feature is the shape of the sheet when viewed from above. How geometry affects sheet rigidity and deflection, however, is not simple to guess. To illustrate this, consider rectangular and triangular sheets of the same length, width, and imposed curvature. By modeling the uncurved portion of the sheet as a cantilever beam, the vertical displacement of the tip can be approximated as  $d=-C\rho g (L-L_p)^4/(E h^2)$, where $C=3/2$ for a rectangle and $C=1/2$ for a triangle \cite{supplement}. The deflection is reduced for a triangular sheet on account of the reduced mass. However, the penetration of the curvature from the boundary, $L_p$, is also diminished. For example, as we shall show later, the persistence length in the curvature-dominated mode is approximately  $L_p^0\sim a/\sqrt{70hR}(a-9 \Delta_p/10)$, for the decrease in sheet width $\Delta_p\ll1$,  recovering the results of Ref.~\cite{btqf14} for a rectangle ($\Delta_p=0$)  and providing a decrease in persistence length for a triangle ($\Delta_p=a L_p^0/L$) \cite{supplement}. These two  effects are in direct competition, prompting the questions of which shape will deflect more and how other sheet geometries affect rigidity.

In this paper, we study the role of material shape on curvature-induced rigidity by combining experiments, finite element simulations, and reduced-order models. The finite element simulations assume a three-dimensional hyperelastic isotropic material \cite{Odgen84}, while the reduced-order model incorporates plate theory \cite{fhmp16} and allows for large longitudinal deflections, akin to a one-dimensional ribbon model \cite{da15, an21, lssm21}. The transition from curvature-dominated to gravity-dominated deformation modes described above is probed, in particular its dependence on the geometry of the thin elastic media. Non-monotonic deflections under gravity are observed for both rectangular and triangular sheets with increasing length. More noticeably, the transition between deformation modes is found to be bistable and hysteretic for a range of sheet lengths and boundary curvatures, $L_0^*\leq L\leq L_1^*$, a range which steadily closes as the sheet is made more triangular.

\section{Experiments}

\begin{figure}[htbp]
\includegraphics[width=0.95\linewidth]{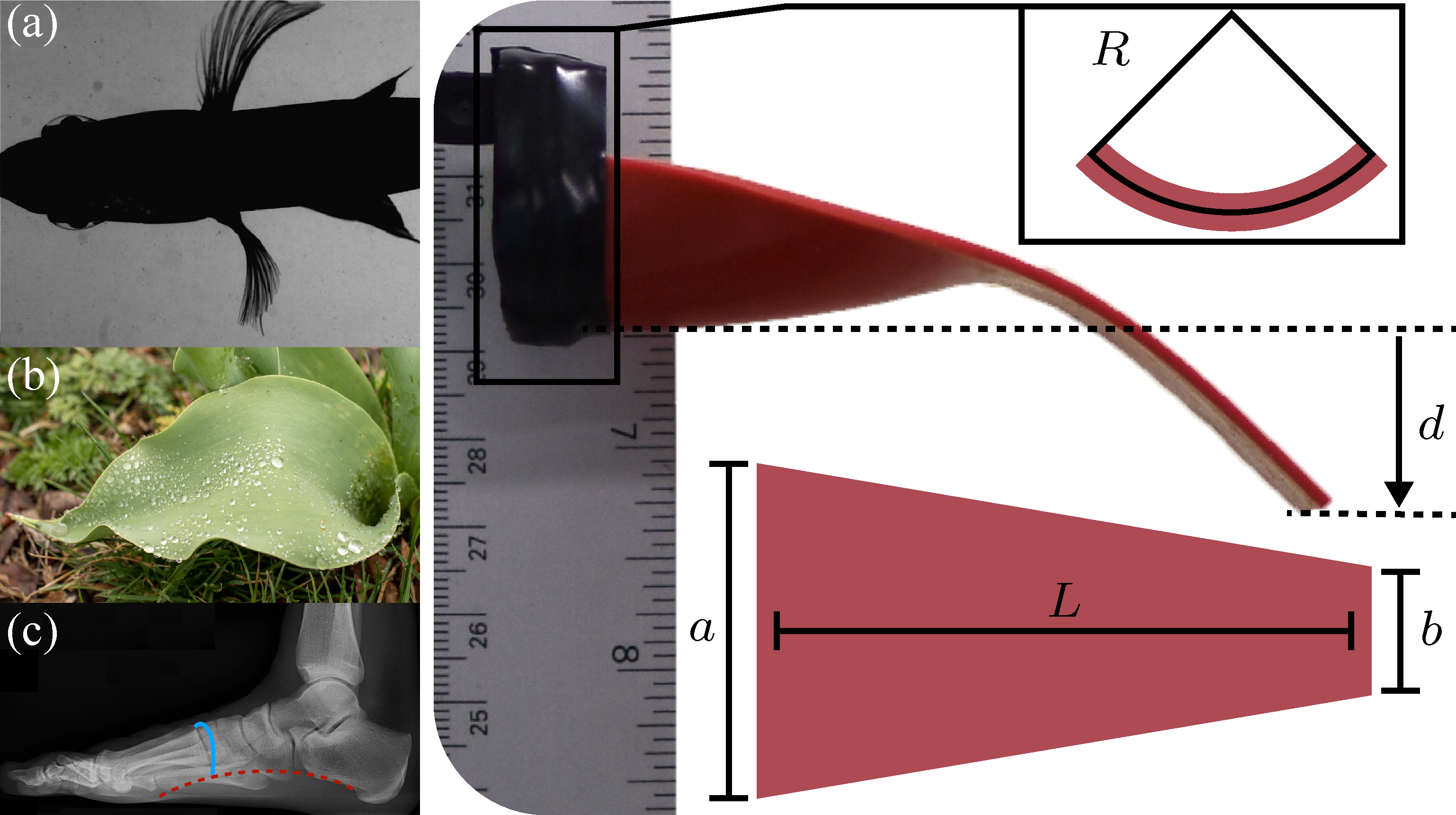}
\caption{Curvature-induced rigidity appears in nature in such disparate systems as: (a) tulip leaves [photograph by H. Mao]; (b) in the human foot (the transverse and longitudinal arches are shown as solid and dashed curves, respectively) [X-ray by M. Häggström, public domain]; (c) fish fins [reproduced with permission from Ref.~\cite{nybvm17}]. (d) Experimental setup. Silicone sheets of thickness $h$, length $L$, base width $a$, and tip width $b$ were clamped at their base upon rigid arcs of radius $R$. The signed tip deflection, $d$, measures the vertical distance between the lowest points of the clamped and free edges.}
\label{Figure1}
\end{figure}

To explore the role of shape on curvature-induced rigidity, thin sheets of commercially available Shore 30A silicone rubber of density $\rho = \SI{1250}{\kilo\gram/\meter^3}$ and thickness $h=\SI{1.5875}{\milli\meter}$ were cut in the shape of a rectangle or a triangle with base width $a = \SI{5}{\centi\meter}$, tip widths $b=a$ (rectangles) and $b=0$ (triangles), and lengths ranging from $L=\SIrange{3}{14}{\centi\meter}$. These were then clamped at their base upon rigid arcs of curvature ranging from $1/R=0$ to $\SI{1}{\per\centi\meter}$, and left free to deform under gravity until the material was fully relaxed (see Fig.~\myref{Figure1}{d}). 

\begin{figure}[htbp]
\includegraphics[width=0.95\linewidth]{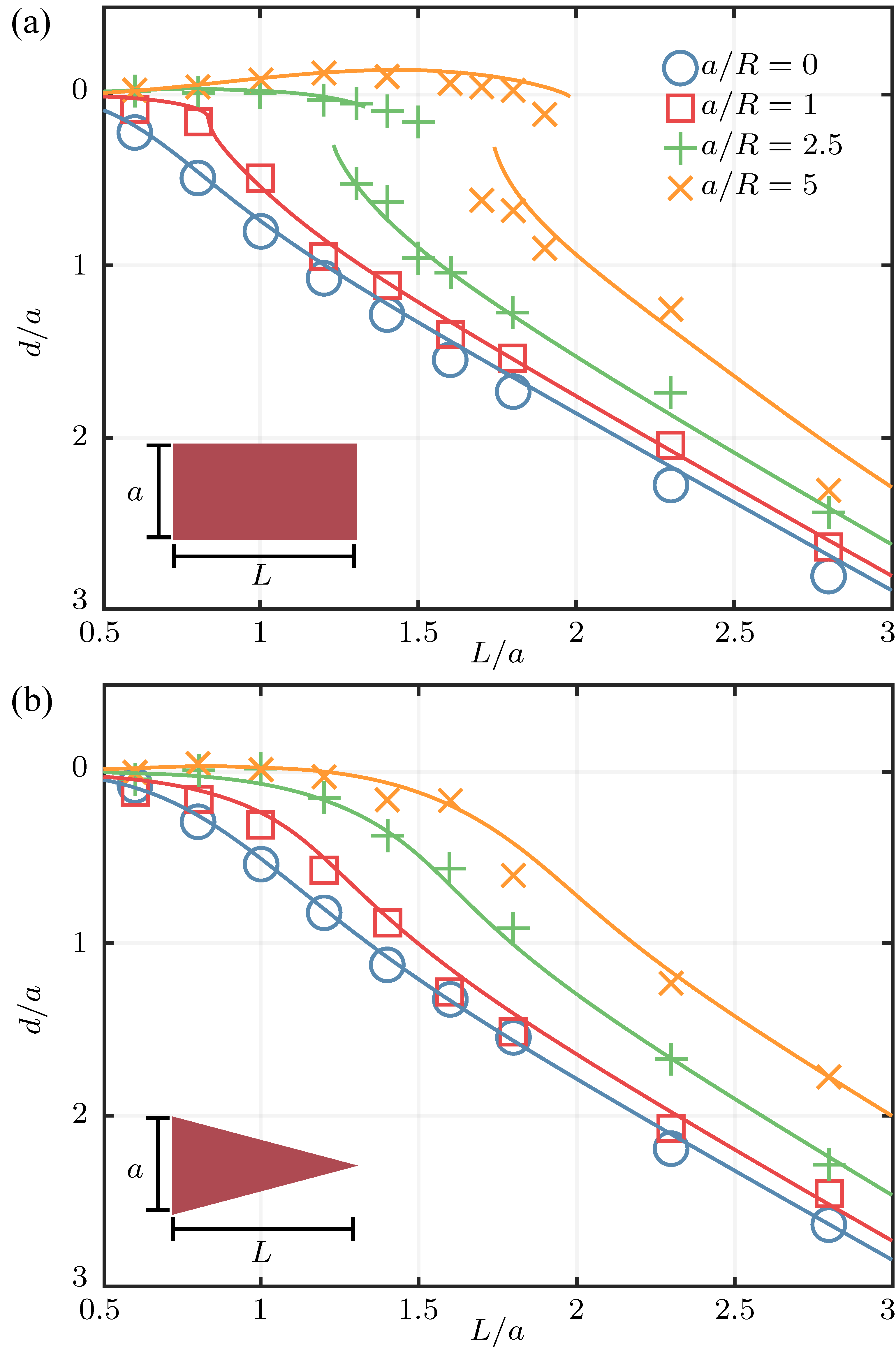}
\caption{The downward tip deflection $d$ as a function of the ratio of sheet length $L$ to base width $a$ from experimental measurements (symbols) and the results of finite element simulations (curves). Different symbols/colors correspond to different clamped curvatures. Curvature-dominated and gravity-dominated configurations, as well as a region of bistability, are observed for rectangular sheets (a). Triangular sheets (b) can deflect more or less than rectangular sheets depending on their length, and do not appear to be bistable.}
\label{Figure2}
\end{figure}

The most apparent measurable quantity is the vertical tip displacement, or deflection, $d$, defined as the signed vertical distance from the lowest point of the clamped edge to that of the free edge. The tip displacement is shown as a function of the relative length, $L/a$, for rectangular ($b/a=1$) and triangular ($b/a=0$) sheets in Fig.~\myref{Figure2}{a} and Fig.~\myref{Figure2}{b}, respectively. A range of clamping curvatures were considered. For zero clamping curvature ($a/R=0$), the tip dropped steadily with increasing sheet length, as expected for a classical cantilever \cite{Wang86,hko08}. Increasing the clamping curvature monotonically decreased the tip deflection for all lengths considered. For large clamping curvatures but small sheet lengths, the tip was observed to deflect upwards against gravity ($d<0$), consistent with previous studies \cite{tbnv19}: the tip is elevated to a position closer to the vertical center of mass, which reduces stretching along the sheet edges.

\begin{figure}[thbp]
\includegraphics[width=0.46\textwidth]{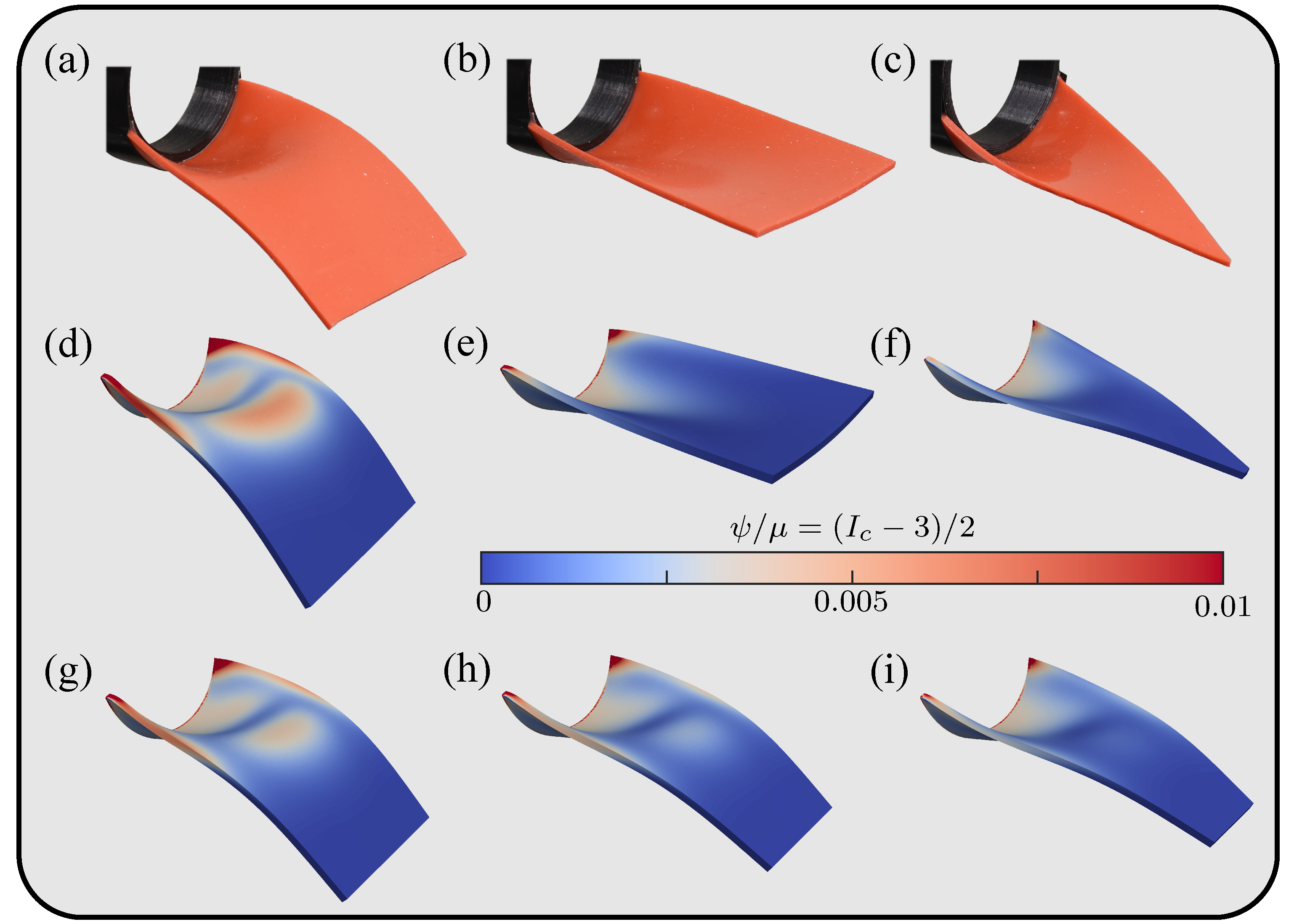}
\caption{A comparison between experimental results and finite element simulations, with base width $a = \SI{5}{\centi\meter}$, clamping curvature $a/R = 2.5$, and sheet length $L/a = 1.3$. \textbf{a--c}~Photographs from experiments: a rectangular sheet in the gravity-dominated mode (a);~the same sheet in the curvature-dominated mode (b); and a triangular sheet in the (sole) gravity-dominated mode (c). \textbf{d--e}: Simulated sheets corresponding to those in \textbf{a--c}, colored by the scaled elastic potential energy $\psi/\mu$. \textbf{g--i}:~ Simulated trapezoidal sheets with shape factors $b/a=0.8$ (g), $0.6$ (h), and $0.4$ (i). The interior saddle-point and localized stretching steadily vanishes as the shape factor $b/a$ decreases, resulting in a smaller tip deflection.}
\label{Figure3}
\end{figure}

Also observed in cases with substantial clamping curvatures ($a/R\gtrsim 1$), at least for rectangular sheets, were regions of bistability, $L_0^*\leq L\leq L_1^*$, where the sheet has two distinct displacement modes. In these ranges, applying a temporary gentle force could change the configuration from one apparently stable state to the other. Figure~\myref{Figure3}{a-c} shows photographs of the two modes for clamping curvature $R/a = 2.5$ and length $L/a = 1.3$. The first mode shows a relaxation of the transverse curvature near the base, beyond which the sheet appears flat as it dips downward due to gravity. The second mode exhibits curvature throughout the length, and no significant tip displacement is observed. Distinct regions of curvature-dominated, gravity-dominated, and commensurate deformations in rectangular sheets were studied by Taffetani et al.~\cite{tbnv19}, who also noted anecdotally the presence of bistability.

\section{Finite-element simulations}

To gain a more detailed view of the elastic deformations in the system, we performed three-dimensional numerical simulations. The sheet is modeled as an isotropic, neo-Hookean material with zero spontaneous curvature \cite{hko08}. The rubber is assumed to be incompressible (i.e., Poisson's ratio $\nu = 0.5$) with Young's modulus $E = \SI{4.5e5}{\pascal}$, a value obtained by comparison with experiments at zero clamping curvature.

The material position and displacement in a flat (undeformed) reference frame are denoted by $\bm{X}$ and $\bm{U}(\bm{X})$, respectively. The elastic energy density is given by $\psi = \mu(I_C-3)/2$ \cite{Odgen84}, where $\mu = E/[2(1+\nu)]$ is the shear modulus and $I_C \coloneqq \tr(\mathbf{C})$, with $\mathbf{C}\coloneqq\mathbf{F}^T\mathbf{F}$ the right Cauchy--Green strain tensor, $\mathbf{F}\coloneqq\mathbf{I}+\nabla \bm{U}^T$ the deformation gradient tensor, and $\mathbf{I}$ the identity matrix. The total energy combines the elastic and gravitational potential energy, $\mathcal{E}(\bm{U})\coloneqq \int_\Omega \psi(\bm{U})+\rho g \hat{\bm{e}}_3\cdot \bm{U}\,\dd V$, where $\rho$ is the material density, $g$ is the gravitational acceleration, $\hat{\bm{e}}_3$ is the vertical unit vector, and $\dd V$ is an infinitesimal volume element in the reference domain $\Omega$. 

The system is made dimensionless by scaling lengths upon the base width $a$ and the energy density upon $\rho g a$. The dimensionless parameters which then govern the equilibrium shape are the relative material length, $L/a$; thickness $h/a$; clamping curvature $a/R$; shape factor (tip-to-base length ratio) $b/a$; and the dimensionless (Young's) modulus, $E/(\rho g a)$. Adjusting the dimensionless modulus alters the balance between elastic and gravitational forces: a larger dimensionless modulus corresponds to a stiffer or less dense sheet. In this study, $E/(\rho g a )=735$ was fixed to match the material used in the experiments.

Energy minimization may be recast as a variational problem. Writing $\Pi(\bm{u}) \coloneqq \mathcal{E}(\bm{U}) - \displaystyle\int_\Omega p(J-1)\ \dd V$ and $\bm{u}\coloneqq(\bm{U},p)$, we seek $\bm{u}\in (H^1(\Omega),L^2(\Omega))$ such that $\dd\Pi(\bm{u} + \varepsilon \bm{v})/\dd\varepsilon\rvert_{\varepsilon = 0}=0$ for all $\bm{v}$ in the same space of functions. Here, we have introduced $J\coloneqq\det(\mathbf{F})$, the determinant of the deformation gradient tensor, and the pressure $p$, a Lagrange multiplier that enforces incompressibility ($J=1$). The displacement of the material is enforced along the clamped surface by prescribing $\bm{U}$ there: the midline is constrained to lie along a circular arc, with lines initially normal to the mid-line on that material face remaining normal and unstretched \cite{supplement}.

The variational problem is solved in a finite element framework using quadratic Lagrange elements \cite{FEniCS15}. The relative material thickness is fixed to align with the experiments, $h/a = 0.03$, and the reference frame is discretized using a tetrahedral mesh with maximum element diameter $\approx 0.03$. The displacement $\bm{U}$ and pressure $p$ are represented as mesh nodal values, which are found by Newton--Raphson iteration. Convergence studies confirmed the predicted second-order accuracy of the tip displacement in the triangle diameter $h$, and higher spatial resolution did not change the results by more than $1\%$~\cite{supplement}. Initial states used in the relaxation were extrapolated from solutions using either smaller gravitational load or smaller clamping curvature, allowing for different configurations to be recovered in the bistable region. 

\subsection{Rectangular and triangular sheets}

Returning to Fig.~\ref{Figure2}, the scaled tip deflections found using simulations are included as solid curves. The experimental and simulated results are generally in good agreement, even before, after, and through the regions of bistability. For zero clamping curvature ($a/R = 0$), the tip deflection increases smoothly with increasing sheet length, consistent with the experimental results. For rectangular sheets with a small clamping curvature ($a/R = 1$), the continuity of the tip deflection with length, or lack thereof, was not immediately clear in the experiments. The simulations, however, suggest a continuous but precipitous increase in the deflection near $L/a=0.8$. By further increasing the clamping curvature, bistability in this region of sheet lengths is observed, as shown by the two branches of the curves corresponding to $a/R = 2.5$ and $a/R = 5$ in Fig.~\myref{Figure2}{a}. For triangular sheets, simulations suggest a smooth change in tip displacement as a function of the sheet length for all clamped curvatures considered, and no sudden transitions were found.

Figure~\myref{Figure3}{d--f} shows three computed sheet profiles colored by the scaled elastic energy, $\psi/\mu=(I_C-3)/2$, revealing good agreement with the respective photographs from corresponding experiments (Fig.~\myref{Figure3}{a--c}). When the rectangular sheet is in the gravity-dominated mode (d), the deformation is dominated by inexpensive transverse bending (with energy scaling with $h^3$) in both the highly curved region near the center of the clamped base and the flat region near the tip. The region connecting the two, however, has non-trivial Gaussian curvature, which in accordance with the \textit{Theorema Egregium} must come with more costly stretching of the material (scaling with $h$) there. This effect is related to the snap-through buckling of a sheet with uniform transverse curvature, i.e, the ``tape-spring'' problem \cite{Mansfield73,sp99,bc10}, where the total energy functional is non-convex \cite{James81, tbnv19}. In the curvature-dominated mode (e), the rectangular sheet has not buckled, and the elastic energy is only substantial in a boundary layer near the corners of the clamped base, due to a disparity in boundary conditions (i.e, a `Saint-Venant' region \cite{hko08}).

Since triangular sheets appear to only have one stable configuration, there is no interior region of large localized elastic energy in Fig.~\myref{Figure3}{f}. The stored energy is instead largest near the boundaries, analogous to the curvature-dominated mode for the rectangular sheet.

\subsection{Trapezoidal sheets}

The apparent buckling of rectangular sheets, but not triangular sheets, motivated a study  of intermediate shapes: trapezoids with shape factors $b/a\in(0,1)$. Example results are shown in Fig.~\myref{Figure3}{g--i}. The interior saddle-point and localized stretching steadily vanishes as the shape factor $b/a$ decreases, resulting in a smaller tip deflection.

\begin{figure}[htbp]
\centering
\includegraphics[width=\linewidth]{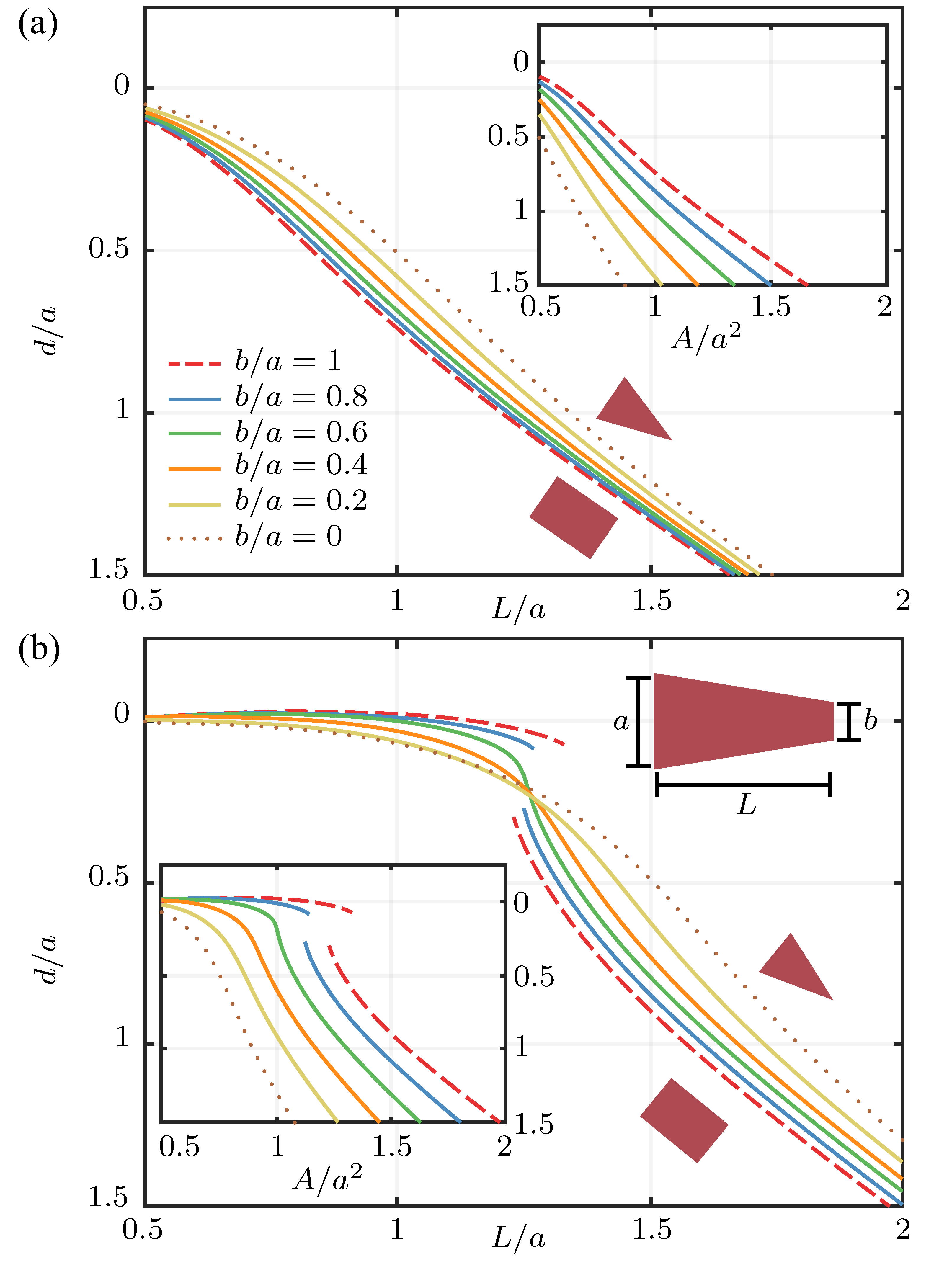}
\caption{The tip deflection, $d/a$, from finite element simulations, as function of the sheet length, $L/a$, for a selection of different shape factors, $b/a$. Two clamping curvatures are considered: $a/R = 0$ (a) and $a/R = 2.5$ (b). Insets: tip deflections as a function of the surface area, $A/a^2$. With no clamping curvature, rectangular sheets deflect more than tapered sheets for fixed length, but the opposite is true for a fixed surface area. For larger clamping curvatures and lengths fixed, triangular sheets deflect more than rectangular sheets below the pop-up length $L_0^*$, and less for lengths above the standout length $L_1^*$. With area instead fixed, sheet deflection is always greater for tapered sheets than that of rectangular sheets.}
\label{Figure4}
\end{figure}

Figure~\myref{Figure4}{a} shows the computed tip displacements as a function of length for a variety of shape factors and zero clamping curvature ($a/R=0$), revealing smooth behavior. For a fixed length, the tip deflection increases with the shape factor due to the larger gravitational load, eventually recovering the classical heavy elastica (i.e., a rectangular sheet, $b/a=1$); but, there is otherwise only a small quantitative dependence on the geometry.

For a large clamping curvatures, however, the behavior is more dramatic (see Fig.~\myref{Figure4}{b} for $a/R = 2.5$, for example). The smooth relationship between the tip deflection and the sheet length, as observed for triangular sheets, begins to sharpen for tapered trapezoidal sheets near $L/a \approx 1.25$. The change becomes extreme at approximately $b/a\approx 0.7$, and the relation becomes multi-valued for yet larger shape factors $b/a$, with a physical saddle point and localized interior stretching coming back into view (Fig.~\myref{Figure3}{g}).  

For a sufficiently short length and large imposed curvature, the sheets do not buckle and the tip displacement is larger for more tapered sheets (smaller $b/a$). For a large fixed length, the tip displacement is instead greater for less tapered sheets (larger $b/a$, see Fig.~\ref{Figure4}), owing to their larger gravitational load. We also take a different view, in which the weight is fixed instead of the length. The two insets in Fig.~\ref{Figure4} show tip displacements as a function of the (scaled) projected surface area, $A/a^2$, for different geometries. More rectangular sheets now deflect less for a fixed area, regardless of the curvature. This reverses the relation observed for a large fixed length, owing to the shorter length of rectangular sheets when compared to tapered sheets of the same weight, allowing the curvature and elastic stresses near the clamped base to play a greater role.

To achieve a deeper understanding of how curvature and geometry affect tip deflection and bistability, we now turn to reduced-order models.

\section{Reduced-order modeling}

\subsection{Persistence length}

The standout length, $L_1^*$, (and hence pop-up length, $L_0^*$) of a curved sheet is typically close to the length at which the curvature imposed by the clamped boundary condition vanishes, i.e., the persistence length $L_p$. In the simplest case, where a rectangular sheet is curved without gravity, the (dimensional) persistence length was found in Ref.~\cite{btqf14} to be given by $L_p = a^2/\sqrt{70hR}$. This approximation is obtained by considering a two-dimensional isotropic elastic plate with a uniform thickness $h$, bending stiffness $D = Eh^3/[12(1-\nu^2)]$, and stretching stiffness $Y = Eh/(1-\nu^2)$ \cite{mansfield89}. The  flat (undeformed) plate is parameterized longitudinally by $s\in[0,L]$ and transversely by $t\in[-a/2,a/2]$. Furthermore, the approximation assumes that points on the curved (deformed) plate (i.e., $0<s<L_p$) can be written in terms of the unit Cartesian basis vectors, $\ee_1$, $\ee_2$, and $\ee_3$, as
\begin{equation}\label{eq:ansatz1}
    \bm{x}(s,t) = \left[s+ U(s) \right]\ee_1+ t \ee_2+ \left[W(s)+c(s)t^2/2\right]\ee_3,
\end{equation}
where $U(s)$ and $W(s)$ are small displacements uniform in each cross-section and $c(s)$ is the lateral curvature. The energy is then dominated by the longitudinal stretching, with strain $E_{11}=U'(s)+(W'(s)+c'(s)t^2/2)^2/2$, and lateral bending, with  strain $B_{11}=c(s)$:
\begin{equation}
    \mathcal{E}\approx\int_0^{L_p}\int_{-a/2}^{a/2}\frac{Y}{2}E_{11}^2 + \frac{D}{2}B_{22}^2\,\dd t\,\dd s.
\end{equation}
Varying this energy with respect to $U(s)$, $W(s)$, and $c(s)$, leads to the persistence length above, $L_p=a^2/\sqrt{70h R}$.

Following the same argument, but replacing the constant width $a$ with a varying width,
\begin{equation}
    p(s)=a+\frac{b-a}{L}s,
\end{equation}
(i.e., $t\in[-p(s)/2,p(s)/2]$) the behavior of the triangular and trapezoidal sheets can also be obtained (see ~\cite{supplement}). In particular, for a small change in sheet width (i.e., $\Delta_p\coloneqq a-p(L_p)\ll a$), we find that
\begin{equation}\label{eq:persistence}
    L_p=\frac{a}{\sqrt{70hR}}\left(a-\frac{9}{10}\Delta_p\right)+\mathcal{O}(\Delta_p^2),
\end{equation}
suggesting a decrease in the persistence length as the tip of the sheet shrinks. This qualitatively agrees with the observation that highly tapered sheets deflect more than rectangular sheets for a fixed small length (i.e., for sheets in the curvature-dominated mode). For long sheets, this decrease in persistence length has a negligible effect on the deflection due to gravity. 

\subsection{Ribbon-like theory}

As a next step we seek a varying-width elastic plate theory, allowing for large longitudinal rotations and small lateral curvatures, similar to recent ribbon \cite{fhmp18, pt19} and rod \cite{hm06, gbcb12, la20} theories. The aim is to describe the material configuration using the cross-section center of mass $\bm{r}(s)$ and its tangent angle $\phi(s)$, as in rod models \cite{al21, kal23}, coupled to equations representing local stretching and bending moments. In place of \eqref{eq:ansatz1},  points on the deformed plate are  written as
\begin{equation}
\bm{x}(s,t) = \bm{r}(s) +u(s, t)\bm{d}_1(s)+  [t + v(s, t)] \bm{d}_2 + w(s, t)\bm{d}_3(s),
\end{equation}
where $u$, $v$, and $w$ are the displacements in their respective directions $\bm{d}_1(s) =\cos\phi(s)\hat{\bm{e}}_1+\sin\phi(s)\hat{\bm{e}}_3$, $\bm{d}_2 =\hat{\bm{e}}_2$, and $\bm{d}_3(s) = -\sin\phi(s)\hat{\bm{e}}_1+\cos\phi(s)\hat{\bm{e}}_3$. Here, $\bm{d}_1\propto \bm{r}'(s)$ is the unit tangent vector pointing along the cross-sectional centers of mass (which is generally outside the material), $\bm{d}_2$ points transversely across the sheet, and $\bm{d}_3=\bm{d}_1\times\bm{d}_2$ is normal to the centerline (but not necessarily normal to the sheet itself) \cite{supplement}.

With the deformation defined above, components of the plate's stretching and bending stress tensors are given by $T_{ij}=Y[(1-\nu)E_{ij}+\nu E_{kk} \delta_{ij}]$ and $M_{ij}=D[(1-\nu)B_{ij}+\nu B_{kk} \delta_{ij}]$, respectively, for  $i,j\in\{s,t\}$, Kronecker delta function $\delta_{ij}$,  strain tensors
$E_{ij} = (\partial_i\bm{x}\cdot\partial_j\bm{x}-\delta_{ij})/2$ and $B_{ij} = \partial_i\partial_j\bm{x}\cdot\hat{\bm{n}}$, and unit normal vector of the plate $\hat{\bm{n}}\propto \partial_s\bm{x}\cross \partial_t\bm{x}$. We also introduce the centerline's strain, $\varepsilon(s)$, such that $\bm{r}'(s)=[1+\varepsilon(s)]\bm{d}_1(s)$. 
The total energy then combines the  stretching and bending energy of the plate  with the gravitational potential energy:
\begin{equation}\label{eq:Energy}
    \mathcal{E} =\int_0^L\int_{-p(s)/2}^{p(s)/2} \frac{1}{2} T_{ij}E_{ij}+ \frac{1}{2}M_{ij}B_{ij}+\rho g  h\hat{\bm{e}}_3\cdot\bm{r}\, \dd t\,\dd  s,
\end{equation}
where repeated indices imply summation.

To reduce the system, the linear and nonlinear contributions to the stretching strains from the in-plane and out-of-plate displacements, respectively, are assumed to be of the same order of magnitude as the longitudinal strain (i.e., $\varepsilon\sim u_s\sim v_t \sim w_t^2\sim\ldots$). Furthermore, assuming $Y\varepsilon^2\sim D(\phi')^2$, so that the energetic cost of stretching balances bending, and $\varepsilon\sim a^2(\phi')^2$, a scaling argument  used to develop F\"oppl--von K\'arm\'an plate theory \cite{mansfield89}, we obtain $\varepsilon\sim h^2/a^2$ and $\phi'\sim h/a^2$. The leading-order strains are then
\begin{subequations}\label{eq:strains}
\begin{gather}
    E_{ss} = \varepsilon - \phi'w+\partial_su+(\partial_sw)^2/2,\\ 
    E_{tt} = \partial_tv + (\partial_tw)^2/2,\\   E_{st} =E_{ts} =(\partial_tu + \partial_sv + \partial_tw \partial_sw)/2,\\
    B_{ss} = \phi' + \partial_{ss}w,\quad
    B_{tt} = \partial_{tt}w,\quad   B_{st} =B_{ts} =\partial_{st}w.\subtag{d--f}
\end{gather}
\end{subequations}
(See the supplementary material, \cite{supplement}, and Ref.~\cite{an21} for further details.)

For small clamping curvatures, separate numerical simulations revealed that the leading-order behavior of the out-of-plane deformation, $w$, is quadratic in $t$, while the in-plane deformations, $u$ and $v$, are higher-order in the clamping curvature (at least away from the buckling region). These observations are consistent with the typical shallow cylinder approximations, as well as the scalings above. We are thus led to assume the ansatz
\begin{subequations}\label{eq:ansatz}
\begin{align}
     w(s,t)= &c(s) \left(\frac{t^2}{2}-\frac{p(s)^2}{4!}\right),\quad v(s,t) =-c(s)^2\frac{t^3}{3!},\subtag{a,b}\\
    u(s,t)= &\frac{c(s)[c(s)p(s)^2]'}{4!} \left(\frac{t^2}{2}-\frac{p(s)^2}{4!}\right)\nonumber\\
    &\quad\quad\quad\quad -c(s)c'(s)\left(\frac{t^4}{4!}-\frac{p(s)^4}{16\times 5!}\right),\subtag{c}
\end{align}
\end{subequations}
where $c(s)$ is the transverse curvature and primes denote derivatives with respect to $s$.  Here, $u$ and $v$ are chosen so that the sheet is strain-free in the transverse direction (i.e., $E_{tt}=E_{st}=0$), while the terms constant in $t$  are chosen so that $\bm{r}$ retains its definition as the centerline. Other possible ansatzes for $u$ and $v$ (for example, imposing that the sheet is transversely stress-free) yield similar results.

Combining \eqref{eq:Energy}--\eqref{eq:ansatz} yields an energy dependent on the sheet's geometry, via $\varepsilon(s)$, $c(s)$, and $\theta(s)$, for a given sheet width profile, $p(s)$.  This energy is then minimized by applying a variational principle (see ~\cite{supplement}). Ultimately, the centerline strain $\varepsilon(s)$ is selected by requiring the average longitudinal stress to vanish, $\int_{-p(s)/2}^{p(s)/2} T_{ss}\,\dd t=0$. The curvature $c(s)$ satisfies a fourth-order ordinary differential equation and the slope $\phi(s)$ satisfies a second-order ordinary differential equation, which  recovers the heavy elastica equation (with  a varying cross-sectional area) when $c(s)=0$. This coupled system is provided in Eqs.~(50--52) in the supplementary material~\cite{supplement} and is solved numerically using a fourth-order accurate finite difference scheme, \texttt{bvp4c} in \textsc{Matlab}.

\begin{figure}[thbp]
\includegraphics[width=0.48\textwidth]{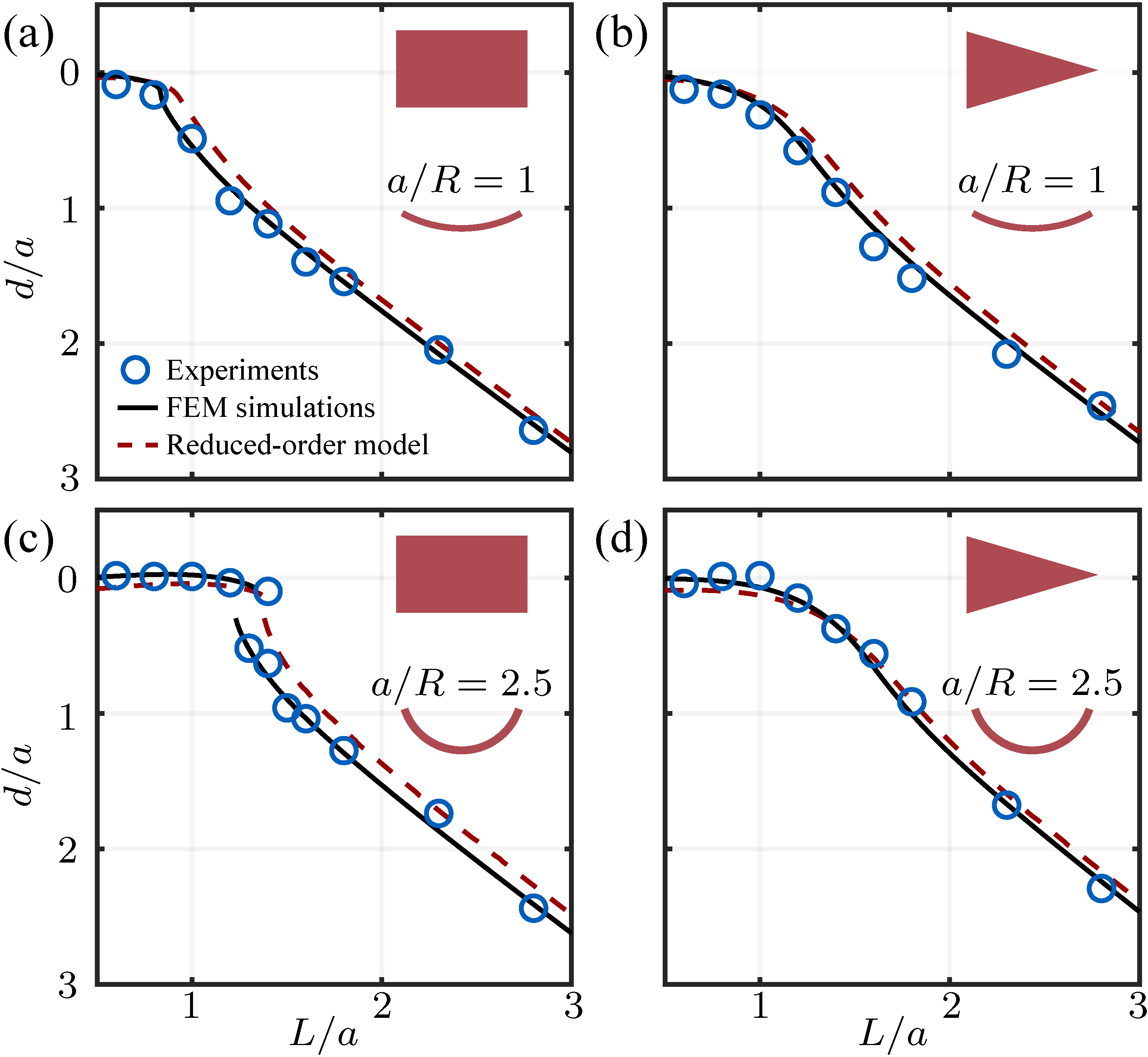}
\caption{Comparison of the tip deflections from experiments (symbols), finite-element simulations (solid curves), and the reduced-order model (dashed curves). (a,b) Rectangular and triangular sheets with smaller clamping curvature, $a/R = 1$. (c,d) Rectangular and triangular sheets with larger clamping curvature, $a/R = 2.5$. The reduced-order model rapidly reproduces the qualitative trends seen using more involved approaches.}
\label{Figure5}
\end{figure}

A comparison of the tip deflections between the experiments, finite element simulations, and the reduced-order approach above is presented in Fig.~\ref{Figure5}, for both rectangular and triangular sheets, and both small and large clamping curvatures. The reduced-order equations capture the qualitative behavior in all cases, and provide quantitatively accurate predictions in most cases. Accuracy is somewhat diminished in the case of large clamping curvatures, which is to be expected due to the shallow-cylinder approximation assumed and the (relatively large) sheet thickness. Nevertheless, the general trend is recovered, including a discontinuous jump in the tip deflection at $L/a=1.38$ (see Fig.~\myref{Figure5}{c}).

\begin{figure}[thbp]
\includegraphics[width=0.98\linewidth]{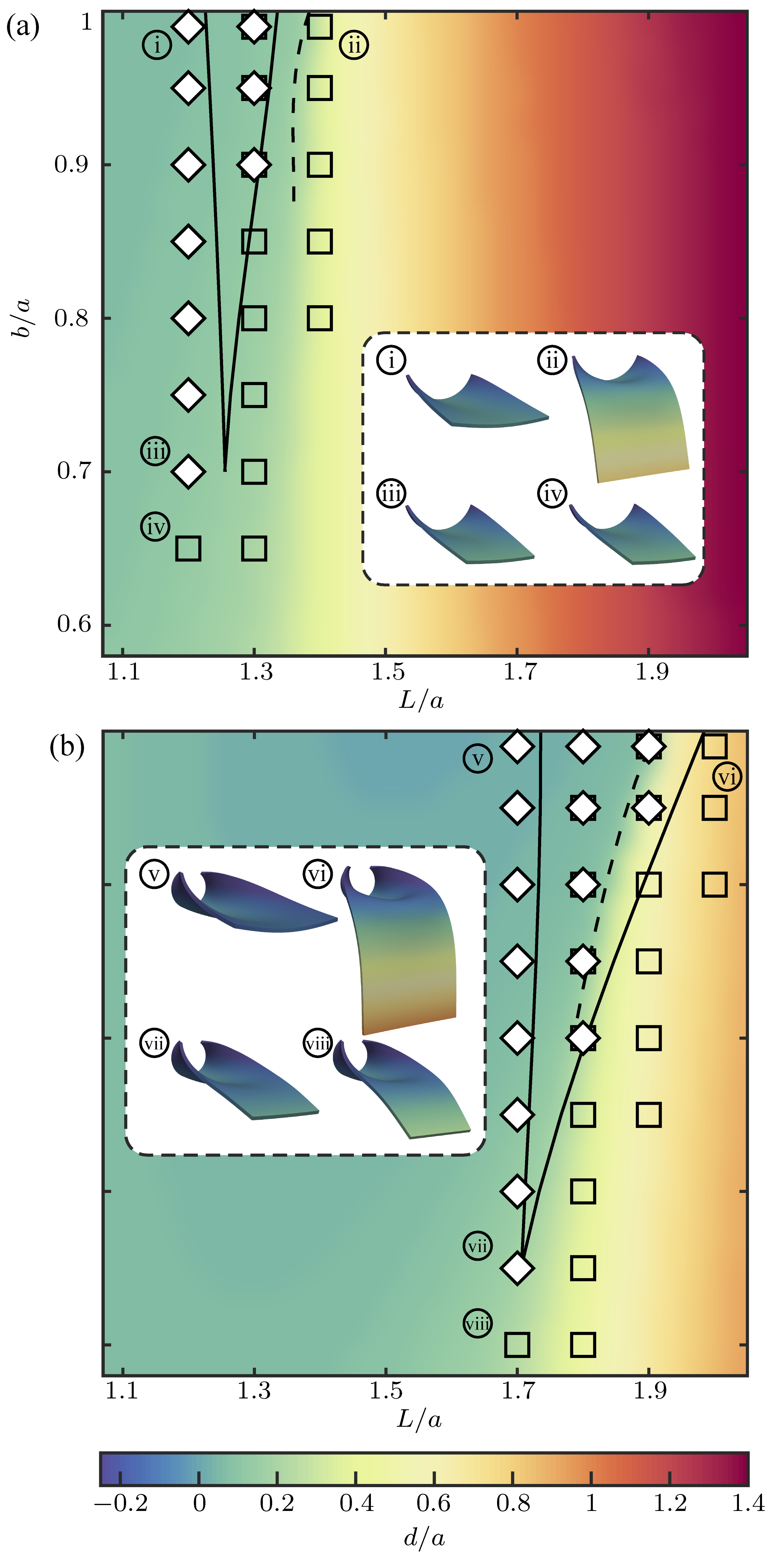}
\caption{Scaled tip displacement, $d/a$, as a function of the scaled sheet length, $L/a$, and shape factor, $b/a$. Values computed using the reduced-order model are shown by the background color, for (a): moderate clamping curvature ($a/R=2.5$); and (b): large clamping curvature ($a/R=5$). Dashed black curves highlight points of rapid discontinuous change. Filled diamonds (open squares) indicate that a curvature-dominated (gravity-dominated) mode was found using finite element simulations at those parameters. Both symbols together indicate bistability, with the solid curves highlighting the boundaries of the bistable region. Inset: physical configurations associated with the labeled points in the parameter space, colored by vertical displacement.}
\label{Figure6}
\end{figure}

Rapid solution of the reduced-order equations allow for a more complete investigation of the relationship between the sheet geometry and the deformation under gravity. Figure~\ref{Figure6} shows color fields representing the tip deflection using the reduced-order model, for moderate clamping curvature ($a/R = 2.5$) and large curvature ($a/R=5$), across a range of sheet lengths and shape factors, $b/a$. Dashed curves indicate sharp transitions between curvature- and gravity-dominated modes for a fixed shape factor and increasing length. This curve shifts rightward for larger clamped curvatures, consistent with the scaling prediction of the persistence length in \eqref{eq:persistence}. This (buckling) discontinuity only exists for a range of shape factors $b/a$, which extends to more tapered shapes at larger clamping curvatures. For even more tapered triangular shapes, the tip deflection is a smooth function of the sheet length.

Overlaid on Figure~\ref{Figure6} are results from the finite element simulations, indicating the observation of curvature-dominated modes (filled diamonds) or gravity-dominated modes (open squares) at each point. The black lines in Fig.~\ref{Figure6} demarcate the region of parameter space for which the finite element simulations reveal bistability, i.e., $L_0^*\leq L\leq L_1^*$ (found using more data points than are shown). With larger clamping curvature, the standout and pop-up lengths for a given shape factor, $L_0^*$ and $L_1^*$, both increase, as was noted qualitatively by the reduced-order model. The critical shape factor $(b/a)^*$, where $L_0^* = L_1^*$, does not appear to be strongly sensitive to the clamping curvature: $L_0^* = L_1^*$ at roughly $(b/a)^* \approx 0.7$ in both  Fig.~\myref{Figure6}{a,b}.

\begin{figure}[thbp]
\includegraphics[width=\linewidth]{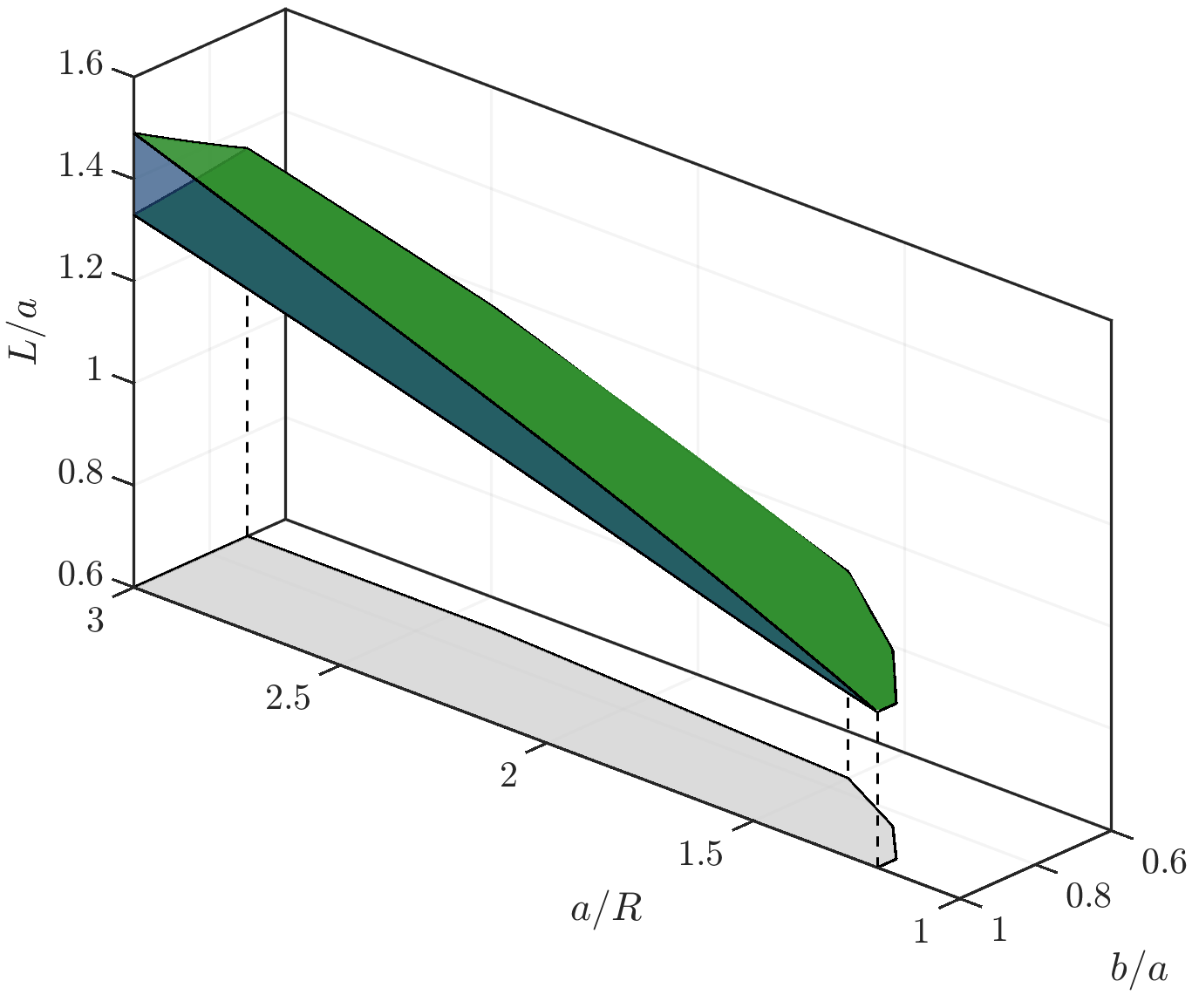}
\caption{Bistability region in the $\{L/a,b/a,a/R\}$ parameter space, obtained by an exhaustive search using finite element simulations. The upper and lower surfaces show the standout lengths ($L^*_1$) and pop-up lengths ($L^*_0$), respectively. Bistability vanishes for sufficiently small clamping curvature or shape factor. Across all clamping curvatures, bistability is only observed for clamping curvatures $a/R \gtrsim 1.2$ and shape factors $b/a\gtrsim 0.6$, for the dimensionless modulus used ($E/(\rho g a )=735$).}
\label{Figure7}
\end{figure}

Finally, we are in position to answer a natural question: for what sheet geometries are there no lengths for which bistability is exhibited, across all clamping curvatures? Figure~\ref{Figure7} shows the bistability region in the $\{L/a,b/a,a/R\}$ parameter space, obtained by an exhaustive search using finite element simulations. The top surface (colored green) shows standout lengths, while the bottom surface (colored blue) indicates pop-up lengths. Bistability is only observed for clamping curvatures $a/R \gtrsim 1.2$, even for rectangular sheets. Both of these surfaces are roughly linear in the three dimensionless parameters, and they intersect along a curve  corresponding to the critical shape factor $(b/a)^*$, now seen to be a function of the clamping curvature. Bistability is most pronounced for large clamping curvatures and rectangular sheets, as was noted previously. But across all clamping curvatures shown, for the relative elastic modulus used ($E/(\rho g a )=735$), there are no trapezoidal sheets with shape factor $b/a\lesssim 0.6$ that exhibit bistability at any length. The bifurcation diagram shown in Fig.~\ref{Figure7} resembles a cusp catastrophe as found in dynamical systems \cite{Strogatz2015}.

Throughout, we have predominately described the buckling and pop-up instabilities by varying the sheet length (moving parallel to the $L/a$ axis in Fig.~\ref{Figure7}); however,  it is also  natural to vary the sheet curvature (moving parallel to the $a/R$ axis). This alternative path leads to the introduction of critical curvatures $1/R_0^*$ and $1/R_1^*$, corresponding to  pop-up (bottom surface) and standout (top surface) instabilities, respectively. One can then move between stable states by curving its end: the tip is elevated when the sheet is curved beyond $1/R_0^*$, and  will remain aloft until the curvature is relaxed below $1/R_1^*$. This may be a more useful approach for applications where the sheet length is fixed.

Although outside the scope of the present study, we expect that increasing $E/(\rho g a)$ (using relatively stiffer sheets) will increase both the standout length $L_0^*$ and the pop-up length $L^*_1$, though their relative changes, and thus the change in the robustness of the bistability, is not simple to predict.

\section{Conclusion}

We have shown through experiments, finite-element simulations, and reduced-order modeling that the shape of a thin elastic material plays a crucial role in curvature-induced rigidity, and hence its deflection under gravity. Rectangular sheets are prone to bistability, while triangular sheets exhibit a more continuous deflection response. Trapezoidal sheets showed a continuous closing of the bistable region as they became more tapered. 
The exploration of Figs.~\ref{Figure6} and \ref{Figure7} also show that bistability may be avoided when transitioning from curvature- to gravity-dominated deformation; for example, by reducing the shape factor, increasing the length, and then increasing the shape factor. This may be of use in many engineering applications where a controlled deflection is required without a discontinuous transition (e.g., buckling).

For small clamping curvatures, whether rectangular or triangular sheets enjoy greater curvature-induced rigidity depends on whether the length is fixed (in which case triangular sheets deflect less) or if the surface area is fixed (in which case rectangular sheets deflect less). For larger clamping curvatures, the competition between the additional weight of the rectangles and the additional lift provided by elasticity was more pronounced: rectangular sheets deflected less before $L_0^*$, the critical pop-up length, but deflected far more for lengths greater than the standout length, $L_1^*$. The numerical simulations and reduced-order model were in close agreement with the experimental findings, allowing for rapid future explorations in related systems. 

As a final application of note, enthusiasts of pizza with non-vanishing thickness should be encouraged. Although triangular slices are generically subject to a gravity-induced tip collapse, one or two bites are usually sufficient to allow curvature-induced rigidity to extend through the remaining roughly trapezoidal material (a bite is a motion through parameter space up and to the left in Fig.~\ref{Figure6}). In addition, savvy diners exploring the bistable regime would be wise to curve the crust before raising the slice from the plate, and not after, so as to select the tidier branch of solutions.

\textit{Acknowledgements.} Support for this research was provided by the Office of the Vice Chancellor for Research and Graduate Education with funding from the Wisconsin Alumni Research Foundation, and by donations to the AMEP program (Applied Math, Engineering, and Physics) at the University of Wisconsin–Madison.

\bibliographystyle{apsrev4-2}
\bibliography{cited}

\newpage 
\textcolor{white}{.}
\newpage 

\onecolumngrid

\begin{center}
\begin{large}
\textbf{Geometric dependence of curvature-induced rigidity}\\
\textbf{-- supplementary information}\\
\vspace{.2cm}
\end{large}
Hanzhang Mao, Thomas G.\ J.\ Chandler, Mark Han and Saverio E. Spagnolie
\end{center} 

\section{Numerical Method}

We consider an elastic material in the shape of a thin trapezoidal prism, with base width $a$, thickness $h$, length $L$, and tip width $b$ curved at end by a radius of curvature $R$. The material position and displacement in a flat (undeformed) reference frame are denoted $\bm{X}=(x,y,z)$ and $\bm{U}(\bm{X})$, respectively. The  elastic energy density is assumed to be the incompressible Neo-Hookean energy $\psi = \mu(I_C-3)/2$, where $\mu = E/[2(1+\nu)]$ is the shear modulus of the material and $I_C = \tr(\mathbf{C})$, with $\mathbf{C}\coloneqq\mathbf{F}^T\mathbf{F}$ the right Cauchy--Green strain tensor, $\mathbf{F}\coloneqq\mathbf{I}+\nabla \bm{U}^T$ the deformation gradient tensor, and $\mathbf{I}$ the identity matrix \cite{Odgen84}. The total energy combines the elastic energy, $\mathcal{E}_{\text{elas}} \coloneqq \int_\Omega \psi(\bm{U})\,\dd V$, and gravitational potential energy, $\mathcal{E}_{\text{grav}} \coloneqq \int_\Omega \rho g \hat{\bm{e}}_3\cdot\bm{U}\,\dd V$, where $\rho$ is the material density (per unit volume), $g$ is the gravitational acceleration, $\hat{\bm{e}}_3=(0,0,1)$ is the vertical unit vector, and $\dd V$ is an infinitesimal volume element in the reference domain $\Omega = \{(x,y,z):x\in[0,L],  y\in[-p(x)/2, p(x)/2],  z\in[-h/2,h/2]\}$, for profile $p(x)=a+(b-a)x$. In addition, a Lagrange multiplier $p$ is introduced to enforce incompressibility, $J \coloneqq \det(\mathbf{F}) = 1$. We, thus, seek to minimize 
\begin{equation}
\Pi(\bm{U},p) \coloneqq \mathcal{E}_{\text{elas}}+\mathcal{E}_{\text{grav}}-\int_\Omega p(J - 1)\,\dd V=\int_\Omega \frac{\mu}{2}(I_C-3)+\rho g \hat{\bm{e}}_3\cdot \bm{U} - p(J - 1)\,\dd V.
\end{equation}

Energy minimization is recast as a variational problem. We seek $\bm{u}\coloneqq (\bm{U},p)\in (H^1(\Omega),L^2(\Omega))$ such that $\dd\Pi(\bm{u} + \varepsilon \bm{v})/\dd\varepsilon\rvert_{\varepsilon = 0}=0$ for all $\bm{v}$ in the same space of functions. The clamped boundary is enforced on the surface  $x=0$ by imposing
\begin{equation}
\bm{U}(0,y,z)=\bm{U}_0(y,z) \coloneqq \left[(R-z)\sin\left(y/R\right)-y\right]\hat{\bm{e}}_2+(R-z)\left[1-\cos\left(y/R\right)\right]\hat{\bm{e}}_3,
\end{equation}
where $\hat{\bm{e}}_1=(1,0,0)$, $\hat{\bm{e}}_2=(0,1,0)$ and $\hat{\bm{e}}_3=(0,0,1)$ are the unit Cartesian vectors. The boundary displacement $\bm{U}_0$ forces the mid-surface ($z=0$) to deform into an arc, with straight lines normal to the mid-surface  remaining unstretched and normal.

\begin{figure}[ht]
\centering
\includegraphics[width=0.45\textwidth]{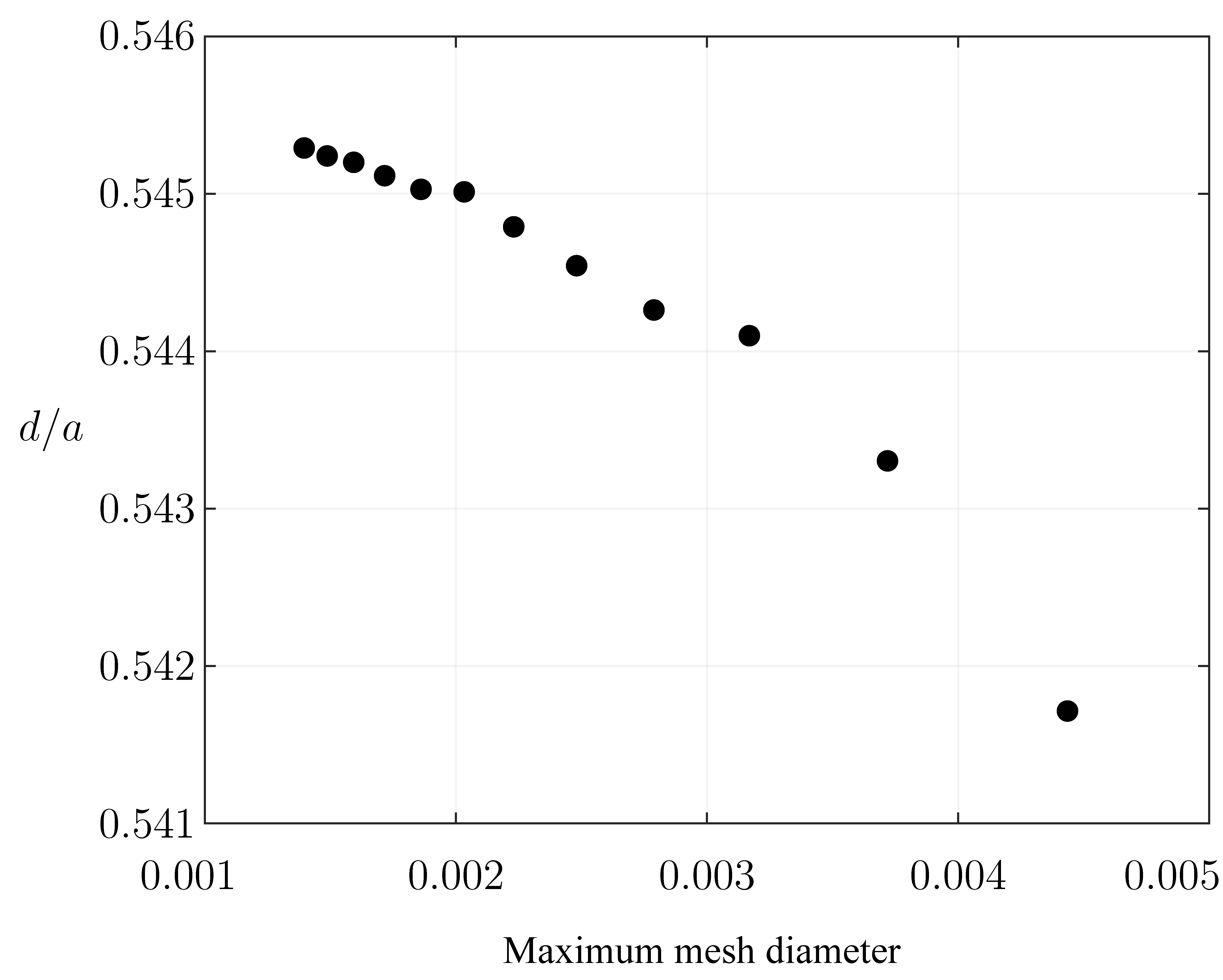}
\caption{Convergence test. The scaled tip deflection $d/a$ is shown as a function of largest mesh diameter for a sheet with $L/a = 1.8$ and $a/R = 5$. With maximum mesh diameter $0.003$, the tip deflection is within $1\%$ of its projected exact value.}
\label{FigureS1}
\end{figure}

The variational problem above was solved with finite elements using quadratic Lagrange elements \cite{FEniCS15}. The reference frame was discretized using a tetrahedral mesh with maximum element diameter $\approx 0.03a$. The displacement $\bm{U}$ and pressure $p$ are represented as mesh nodal values, which are found by Newton--Raphson iteration.  The relative material thickness is fixed to align with the experiments, $h/a = 0.03$. Convergence studies confirmed the predicted second-order accuracy of the tip displacement in the triangle diameter, and higher spatial resolution did not change the results by more than $1\%$. We choose a sheet of length $L/a = 1.8$ and clamping curvature $a/R = 5$ as a demonstration in Fig. \ref{FigureS1} due to its relatively large strain and deformation compared to other sheets considered.

To achieve numerical convergence, the gravitational load and imposed curvature is increased iteratively starting at no gravity and no curvature, with $\bm{U}(\bm{X}) = \bm{0}$ and $p(\bm{X})=0$. Each step uses the previous step's solution of $\bm{U}$ and its Lagrange multiplier $p$ as the initial guess for the Newton--Raphson iteration. In the bistable region, increasing gravitational load first will typically result in a gravity-dominated mode, whereas increasing clamping curvature first will typically result in a curvature-dominated mode. In this way we are able to select the stable state to obtain.

Solutions for gravity-dominated deformation of fixed clamping curvature $a/R$ and tip-to-base ratio $b/a$ is iterated starting from a long sheet ($L/a \sim 2$) and decreasing its length until a non-convergence is reached. Solutions for curvature-dominated deformation is obtained in a similar manner.

In both of these two ways of iteration, a non-convergence means that between the two steps the change in $\bm{U}(\bm{X})$ is large, often implying a discontinuous change in deformation mode.

\section{Reduced-order theory: small plate rotations}\label{sec:persistence}

\begin{figure}[ht]
\centering
\includegraphics[width=0.8\textwidth]{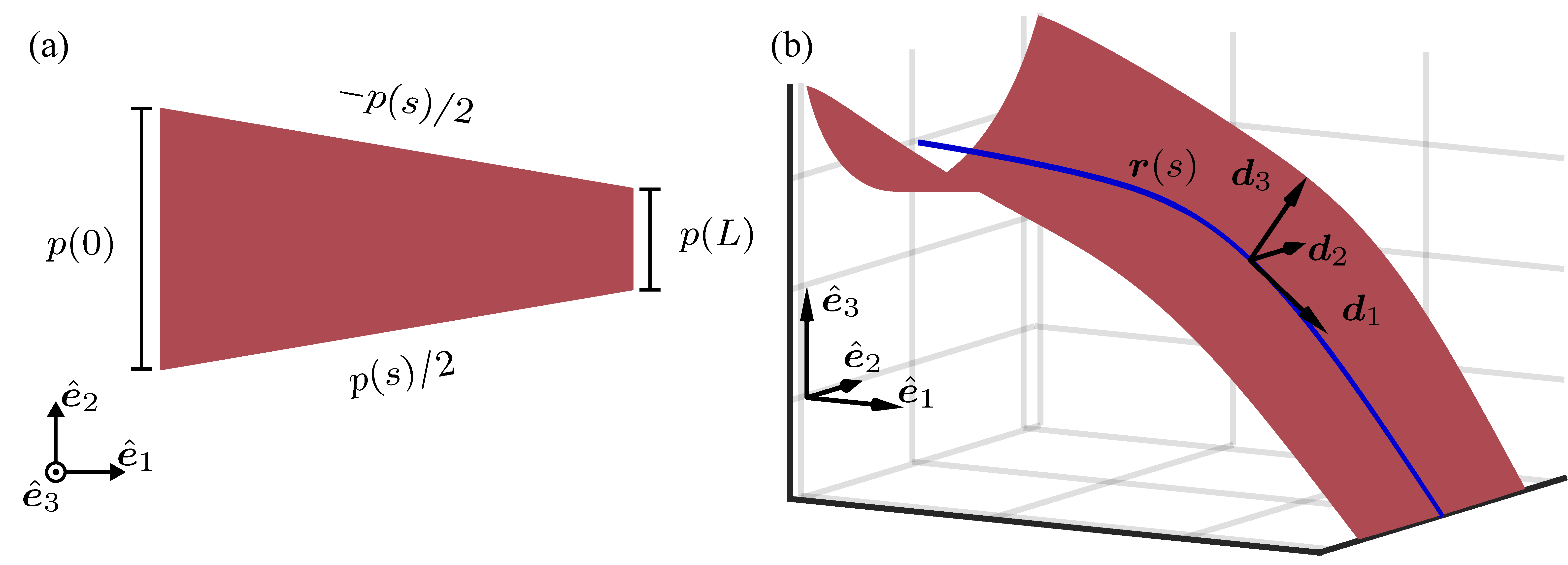}
\caption{(a) Schematic of a sheet in the reference (undeformed) configuration when viewed from above, parameterized by $(s,t)$, with arclength $s\in[0,L]$   and $t\in[-p(s)/2,p(s)/2]$. (b) One-dimensional representation in the reduced-order model. The cross-sectional center of mass, $\bm{r}(s)$ (shown as a blue curve) is not generally internal to the material. An orthogonal frame, $\{\bm{d}_1,\bm{d}_2,\bm{d}_3\}$, includes the tangent vector $\bm{d}_1$ and  normal vector $\bm{d}_3$ (which is not generally normal to the sheet).}
\label{FigureS2}
\end{figure}

Consider a two-dimensional plate with length $L$, uniform thickness $h$, Poisson's ratio $\nu$, stretching modulus $Y= Eh/(1-\nu^2)$, and bending modulus $D=Eh^3/[12(1-\nu^2)]$. The position of the flat (undeformed) plate is 
\begin{equation}\label{eq:undeformed}
    \bm{X}(s,t)=s\,\hat{\bm{e}}_1+t\,\hat{\bm{e}}_2,
\end{equation}
with unit Cartesian basis vectors $\hat{\bm{e}}_1$, $\hat{\bm{e}}_2$, and $\hat{\bm{e}}_3$,  longitudinal arclength  $s\in[0,L]$, and lateral arclength $t\in[-p(s)/2, p(s)/2]$, where $p(s)$ is the width of the plate  a distance $s$ from its clamping edge. This undeformed configuration is shown in Fig.~\myref{FigureS2}{a}. (For example, $p(s)=a+(b-a)s/L$ for the trapezoidal shapes considered in the main text.) Denoting the displacements in the $\hat{\bm{e}}_1$, $\hat{\bm{e}}_2$, and $\hat{\bm{e}}_3$ directions as $U(s,t)$, $V(s,t)$, and $W(s,t)$, respectively, the position of the deformed strip is
\begin{equation}
    \bm{x}(s,t)=\left[s+U(s,t)\right]\,\hat{\bm{e}}_1+\left[t+V(s,t)\right]\,\hat{\bm{e}}_2+W(s,t)\hat{\bm{e}}_3.
\end{equation}

According to geometrically nonlinear plate theory \cite{mansfield89}, for small displacements,  the  membrane (stretching) strains within the plate are
\begin{subequations}
\begin{equation}
    E_{11} =\partial_s U+(\partial_s W)^2/2,\quad E_{22} =\partial_t V+(\partial_t W)^2/2, \quad E_{12} =(\partial_t U+\partial_s V+\partial_t W \partial_s W)/2, \subtag{a--c}
\end{equation}
\end{subequations}
while the  bending (curvature) strains are
\begin{subequations}
\begin{equation}
    B_{11} =\partial_{ss}W,\quad B_{22} =\partial_{tt} W , \quad B_{12} =\partial_{st} W. \subtag{a--c}
\end{equation}
\end{subequations}
Furthermore, the components of the plate's stretching stress tensor and bending stress tensor are 
\begin{subequations}
\begin{equation}
    T_{ij}= Y[(1-\nu)E_{i j}+\nu E_{kk}\delta_{ij}]\quad\text{and}\quad M_{ij}= D[(1-\nu)B_{i j}+\nu B_{kk}\delta_{ij}],\subtag{a,b}
\end{equation}
\end{subequations}
respectively, where $\delta_{ij}$ is the Kronecker delta function and repeated indices denote summation. The elastic   and gravitational potential energy stored within the deformed plate is then given by
\begin{equation}\label{eq:energy_simple}
    \mathcal{E}=\int_0^L \int_{-p(s)/2}^{p(s)/2} \frac{1}{2}T_{ij}E_{ij}+\frac{1}{2}M_{ij}B_{ij}+\rho g h W \,\dd t\,\dd s,
\end{equation}
where $\rho g$ is the weight density (per unit volume).

\subsection{Persistence length in the absence of gravity}\label{sec:flat}

We begin by extending the derivation by Barois \emph{et al.}~\cite{btqf14} (which are also presented by Taffetani \emph{et al.}~\cite{tbnv19})  for the persistence length of a rectangular plate in the absence of gravity, to a plate with width that varies along its length. Following  Barois \emph{et al.}~\cite{btqf14} and Taffetani \emph{et al.}~\cite{tbnv19}, we consider an ansatz  of the form
\begin{subequations}\label{eq:ansatz1sup}
\begin{equation}
    U(s,t) = U_0(s), \quad V(s,t)= 0, \quad W(s,t) = W_0(s)+ c(s)t^2/2, \subtag{a--c}
\end{equation}
\end{subequations}
where $U_0(s)$ and $W_0(s)$ are displacements uniform in each cross-section and $c(s)$ is the lateral curvature. Provided the imposed curvature is sufficiently large, the longitudinal stretching (with strain $E_{11}=U_0'(s)+[W_0'(s)+c'(s)t^2/2]^2/2$) and lateral bending (with strain $B_{22}=c(s)$)  dominate the plate's elastic response, and  so the energy \eqref{eq:energy_simple} is approximated as
\begin{equation}\label{eq:Energyapprox}
    \mathcal{E}\approx \int_0^{L}\int_{-p(s)/2}^{p(s)/2} \frac{Y}{2}E_{11}^2+\frac{D}{2}B_{22}^2\,\dd t\,\dd s = \int_0^{L}\int_{-p(s)/2}^{p(s)/2} \frac{Y}{2}\left(U_0'(s)+\frac{1}{2}W_0'(s)^2+\frac{t^2}{2}W_0'(s)c'(s)+\frac{t^4}{8}c'(s)^2\right)^2+\frac{D}{2}c(s)^2\,\dd t\,\dd s.
\end{equation}
Varying this energy  with respect to $U_0'(s)$ and $W_0'(s)$, results in the ODEs
\begin{equation}\label{eq:U0W0}
U_0'(s) =-\frac{3}{3920}c'(s)^2p(s)^4 \quad \text{and} \quad W_0'(s)=-\frac{3}{56}c'(s)p(s)^2, \subtag{a,b}
\end{equation}
which are subject to  clamping  conditions $U_0(0)=0$ and $W_0(0)=0$. Note that the coefficient of $U_0'(s)$ differs in sign from that found by Taffetani \emph{et al.}~due to them absorbing $W_0'(s)^2/2$ into $U_0'(s)$, however this does not affect the results below, and our coefficents do match those found by Barois \emph{et al.}

Inserting \eqref{eq:U0W0} into \eqref{eq:Energyapprox} and integrating yields the reduced energy
\begin{equation}\label{eq:Energy1}
    \mathcal{E}\approx \int_0^{L} \frac{Y }{140 \times 8!}p(s)^9c'(s)^4+\frac{D}{2}p(s)c(s)^2\,\dd s.
\end{equation}
Varying the curvature $c$ yields the Euler--Langrange equation
\begin{equation}\label{eq:code}
    \frac{Y}{35 \times 8!}\left[p(s)^9c'(s)^3\right]'=Dp(s)c(s).
\end{equation}
which is subject to an imposed curvature at $s=0$ and natural boundary conditions at $s=L$, i.e., $c(0)=1/R$ and $c'(L)=0$, respectively. Solutions to this BVP  yield the evolution of the lateral curvature for $L<L_p$. As $L$ increases, there exists a critical length, $L=L_p$, at which the plate's end curvature vanishes, i.e., $c(L_p)=0$. For $L>L_p$, the sheet is then split in two: the curved region, $0<s<L_p$, in which $c(s)$ satisfies \eqref{eq:code} subject to $c(0)=1/R$, $c(L_p)=0$, and $c'(L_p)=0$,  and the flat region, $L_p<s<L$, in which $c(s)=0$. The persistence length is then found by solving this over-constrained boundary value problem (BVP), which we consider now.

\paragraph{Dimensionless system}
To analyze this problem further, we introduce the dimensionless variables
\begin{equation}
    S  \coloneqq s/L_p, \quad C(S) \coloneqq R c(s),\quad P(S) =  p(s)/a.
\end{equation}
The ODE in \eqref{eq:code} is then
\begin{equation}\label{eq:code2}
\left[P(S)^3C'(S)\right]^2\left[P(S)^3C'(S)\right]'= 8\hat{L}_p^4 P(S)C(S),
\end{equation}
which is subject to
\begin{subequations}\label{eq:bc2}
\begin{equation}
C(0)=1,\quad C(1)=0,\quad C'(1)=0,\subtag{a--c}
\end{equation}
\end{subequations}
for the (unknown) dimensionless persistence length
\begin{equation}
    \hat{L}_p \coloneqq \frac{\sqrt{70 h R}}{a} \frac{L_p}{a}. 
\end{equation}

\paragraph{Rectangular sheet} For a rectangular sheet, $P(S)=1$, multiplying  \eqref{eq:code2} by $C'(S)$ and integrating  yields
\begin{equation}
       C'(S)^4=  16\hat{L}_p^4C(S)^2,
\end{equation}
where we have set the integration to zero after imposing the boundary conditions, \eqref{eq:bc2}. Integrating and applying the boundary conditions again  yields the curvature  and persistence length
\begin{subequations}
\begin{equation}
    C(S)=(1-S)^2 \quad\text{and}\quad \hat{L}_p=1.\subtag{a,b}
\end{equation}
\end{subequations}
Thus, the dimensional persistence length is 
\begin{equation}
    L_p=\frac{a^2}{\sqrt{70hR}},
\end{equation}
which recovers the results found in Ref.~\cite{btqf14}.

\paragraph{Triangular and trapezoidal sheets} For a trapezoidal ($b>0$) and triangular ($b=0$) sheet, $P(S)=1-\hat\Delta_p S$, where $\hat\Delta_p\coloneqq\Delta_p/a$ for the decrease in sheet width $\Delta_p=(a-b)L_p/L$. Here, \eqref{eq:code2} becomes
\begin{equation}\label{eq:code3}
\left[(1-\hat\Delta_p S)^3C'(S)\right]^2\left[(1-\hat\Delta_p S)^3C'(S)\right]'= 8\hat{L}_p^4 (1-\hat\Delta_p S)C(S).
\end{equation}
To our knowledge this can not be integrated explicitly, however progress can be made in the asymptotic limit of $\hat\Delta_p\to 0$ (i.e., $L\gg L_p$), which we show now. As $\hat\Delta_p\to 0$, the rectangular sheet problem is recovered at leading-order; thus, we pose the expansion
\begin{subequations}
\begin{equation}
    C(S)= (1-S)^2+\hat\Delta_p C_1(S) +\Oh(\hat\Delta_p^2) \quad\text{with}\quad \hat{L}_p=1+\hat\Delta_p \hat{L}_{1}+\Oh(\hat\Delta_p^2),
\end{equation}
\end{subequations}
where the corrections, $ C_1(S)$ and $\hat{L}_{1}$, are to be determined now. At $\Oh(\hat\Delta_p)$, \eqref{eq:code3} yields
\begin{equation}
(1-S)^2C_1''(S)-2(1-S)C_1'(S)-2C_1(S)=2(1-S)^2(11S+4\hat{L}_{1}-3),
\end{equation}
subject to $ C_1(0)=C_1(1)=C_1'(1)=0$. This can be solved to give the curvature and persistence length correction: 
\begin{equation}
C_1(S)=\frac{11}{5}S(1-S)^2 \quad\text{and}\quad \hat{L}_1=-9/10.
\end{equation}
Thus, the dimensional persistence length is
\begin{equation}
    L_p=\frac{a}{\sqrt{70hR}}\left(a-\frac{9}{10}\Delta_p\right)+\Oh(\Delta_p^2),
\end{equation}
as $\Delta_p=(a-b)L_p/L\to 0$. Or equivalently, in terms of the shape factor $b/a$, 
\begin{equation}
    L_p\sim \alpha a \left(1-\frac{9 \alpha a}{10 L} \left(1-\frac{b}{a}\right)\right)
\end{equation}
as $(a-b)\alpha/L \to 0$, where $\alpha=a/\sqrt{70 h R}$.

\subsection{Vertical displacement in the absence of curvature}\label{sec:EulerBernoulli}

For plates longer than its persistence length, $L>L_p$, the lateral curvature of the plate is not sufficient to keep it aloft against gravity. Instead, we expect the plate to bend longitudinally, leading to a vertical displacement.  As a first approximation of this, we  assume that gravity only acts on the flat region of the plate ($L_p<s<L$), and  the curved region ($0<s<L_p$) is still governed by the gravity-less problem in \S\ref{sec:flat}.

Here, the small displacements are expected to be uniform in each cross-section, and so, the ansatz \eqref{eq:ansatz1sup} becomes
\begin{subequations}
\begin{equation}
    U(s,t) = -W_1(s)^2/2, \quad V(s,t)= 0, \quad W(s,t) = W_1(s), \subtag{a--c}
\end{equation}
\end{subequations}
where $U$ is chosen so that the longitudinal stretching strain vanishes, $E_{11}=0$.  The only non-zero strain is the longitudinal bending strain $B_{11}=W_0''(s)$, and the energy \eqref{eq:energy_simple} becomes
\begin{equation}
    \mathcal{E}=\int_{L_p}^L\int_{-p(s)/2}^{p(s)/2}\frac{D}{2} B_{11}^2+\rho g h W\,\dd t\, \dd s =\int_{L_p}^L\frac{D}{2} W_1''(s)^2+\rho g h p(s) W_1(s)\,\dd s,
\end{equation}
where we have suitably restricted our attention to $L_p<s<L$.  Varying this energy with respect to $W_1(s)$ yields the Euler--Bernoulli cantilever equation for a thin plate with  varying width $p(s)$,
\begin{equation}
    D\left[p(s)W_1''(s)\right]''+\rho g h p(s)=0,
\end{equation}
which is subject to continuity  at $s=L_p$ and natural boundary conditions at $s=L$,
\begin{equation}
    W_1(L_p)=W_0(L_p),\quad W_1'(L_p)=0,\quad W_1''(L)=0,\quad W_1'''(L)=0.\subtag{a--d}
\end{equation}

For a rectangle ($p(s)=a$) and triangle ($p(s)=a(1-s/L)$) the BVP can be solved exactly to give 
\begin{equation}
    W_1(s)=W_0(L_p)-C\frac{\rho g h }{3D}(s-L_p)^2\left[(s-L_p)^2-4(L-L_p)(s-L_p)+6(L-L_p)^2\right],
\end{equation}
where $C=1/24$ for a triangle and $C=1/8$ for a rectangle. This yields the displacement of the flat plate
\begin{equation}
    W_1(L)-W_0(L_p)=-C\frac{\rho g h }{D}(L-L_p)^4,
\end{equation}
as given in the main text.

\section{Reduced-order theory: large plate rotations}

In this section, we improve the \emph{ad hoc} plate theory used in \S\ref{sec:persistence}, to allow for large plate rotations in the longitudinal direction. This formulation will couple transverse curvature and gravity-driven deflections of the plate using a theory similar to some recent ribbon theories \cite{fhmp18, pt19}.

Consider the same undeformed two-dimensional  plate as described in \S\ref{sec:persistence} with undeformed position given in \eqref{eq:undeformed} and sketched in Fig.~\myref{FigureS2}{a}. In contrast to the previous model, here positions on the  deformed plate, $\bm{x}(s,t)$, are written in terms of the cross-sectional center of mass $\bm{r}(s)=\overline{\bm{x}(s,t)}$, where  $\overline{*}\coloneqq \frac{1}{p}\int_{-p/2}^{p/2}*\, \dd t$ is the cross-sectional average, and it's tangent angle $\phi(s)$. That is
\begin{equation}
\bm{x}(s,t) = \bm{r}(s) +u(s, t)\,\bm{d}_1(s)+  \left[t + v(s, t)\right] \bm{d}_2 + w(s, t)\,\bm{d}_3(s),
\end{equation}
where $u$, $v$,  and $w$ are the displacements in the orthonormal directions
\begin{subequations}\label{eq:orthonormal}
\begin{equation}
    \bm{d}_1(s) =\cos\phi(s)\hat{\bm{e}}_1+\sin\phi(s)\hat{\bm{e}}_3,\quad \bm{d}_2 =\hat{\bm{e}}_2,\quad \bm{d}_3(s) =-\sin\phi(s)\hat{\bm{e}}_1+\cos\phi(s)\hat{\bm{e}}_3,\subtag{a--c}
\end{equation}
\end{subequations}
for the Cartesian basis vectors $\hat{\bm{e}}_1$, $\hat{\bm{e}}_2$, and $\hat{\bm{e}}_3$. Here, $u$, $v$, and $w$ are defined relative to the center of mass with $\overline{u(s,t)}=\overline{v(s,t)}=\overline{w(s,t)}=0$. The deformed configuration is shown in Fig.~\myref{FigureS2}{b}.

With the above deformations, the components of the stretching stress tensor, $T_{ij}$, and bending stress tensor, $M_{ij}$ are
\begin{subequations}
\begin{equation}
    T_{ij}= Y[(1-\nu)E_{i j}+\nu E_{kk}\delta_{ij}]\quad\text{and}\quad M_{ij}= D[(1-\nu)B_{i j}+\nu B_{kk}\delta_{ij}],\subtag{a,b}
\end{equation}
\end{subequations}
for the stretching modulus $Y= Eh/(1-\nu^2)$, bending modulus $D=Eh^3/[12(1-\nu^2)]$, and components of the stretching strain tensor, $E_{ij}$,  and bending strain tensors, $B_{ij}$, given by
\begin{subequations}
\begin{equation}
    E_{ij}= (\partial_i\bm{x}\cdot\partial_j\bm{x}-\delta_{ij})/2 \quad\text{and}\quad B_{ij}= \partial_i\partial_j\bm{x}\cdot\hat{\bm{n}},\subtag{a,b}
\end{equation}
\end{subequations}
for $i,j\in\{s,t\}$, Kronecker delta function $\delta_{ij}$, and unit normal vector $\hat{\bm{n}} =(\partial_{s}\bm{x}\times \partial_{t}\bm{x})/\lvert\partial_{s}\bm{x}\times \partial_{t}\bm{x}\rvert$. We also introduce the strain  of the center of mass, $\varepsilon(s)$, such that 
\begin{equation}
    \bm{r}'(s) = [1+\varepsilon(s)]\bm{d}_1(s).
\end{equation}

The total potential energy of the plate  combines the  stretching,  bending, and gravitational potential energies:
\begin{equation}\label{eq:energy}
    \mathcal{E}=\int_0^L \int_{-p(s)/2}^{p(s)/2} \frac{1}{2}T_{ij}E_{ij}+\frac{1}{2}M_{ij}B_{ij}+\rho g h \hat{\bm{e}}_3\cdot \bm{r}(s)\,\dd t\,\dd s
\end{equation}
 where $\rho g h$ is the weight density (per unit area) of the plate and repeated indices denote summation. 

To reduce this system, we shall assume the scaling discussed in the main text: $s\sim t\sim a$, $\varepsilon\sim H^2$, $\phi'\sim H/a$, $u\sim v\sim a H^2$, and $w\sim a H$,  where $p(0)=a$ is the width at $s=0$ and  $H=h/a$ is the relative thickness. As $H\to 0$, the center of mass and deformed position become 
\begin{subequations}\label{eq:positionasym}
\begin{equation}
    \bm{r}(s)= \int_0^s \bm{d}_1(\xi)\,\dd \xi+\Oh(H^2)\quad\text{and}\quad \bm{x}(s,t) = \bm{r}(s)+ t\,\bm{d}_2+  w(s,t)\bm{d}_3(s)+\Oh(H^2),\subtag{a,b}
\end{equation}
\end{subequations}
respectively, while the stretching and bending strain components are
\begin{subequations}\label{eq:stretchstrains}
\begin{gather}
E_{ss} = \varepsilon(s) - \phi'(s)w+\partial_su+(\partial_sw)^2/2+\Oh(H^3),\quad
    E_{tt} = \partial_tv + (\partial_tw)^2/2+\Oh(H^3),\subtag{a,b}\\
    E_{st} =E_{ts} =(\partial_tu + \partial_sv + \partial_tw \partial_sw)/2+\Oh(H^3),\subtag{c}
    \end{gather}
\end{subequations}
and 
\begin{subequations}\label{eq:bendstrains}
\begin{gather}
    B_{ss} = \phi'(s) + \partial_{ss}w+\Oh(H^2),\quad
    B_{tt} = \partial_{tt}w+\Oh(H^2),\quad   B_{st} =B_{ts} =\partial_{st}w+\Oh(H^2).\subtag{a--c}
    \end{gather}
\end{subequations}
Inserting \eqref{eq:positionasym}--\eqref{eq:bendstrains} into \eqref{eq:energy}, and noting that $D/Y=\Oh(H^2)$, yields the leading-order energy  as a function of the intrinsic equations $\varepsilon(s)$, $\phi(s)$, $u(s,t)$, $v(s,t)$, and $w(s,t)$:
\begin{equation}\label{eq:asymenergy}
    \mathcal{E}\sim\int_0^L\left[ \frac{p(s)}{2}\overline{T_{ij}E_{ij}}+\frac{p(s)}{2}\overline{M_{ij}B_{ij}}+\rho g h p(s)\int_0^s\sin\phi(\xi)\,\dd\xi\right]\,\dd s.
\end{equation}
Before minimizing this energy, we first consider what occurs at the clamping edge at $s=0$.

\paragraph{Clamping conditions}  At $s=0$, we  assume   the plate is clamped horizontally to a circle of curvature of $1/R$. That is:
\begin{subequations}
\begin{equation}
    \bm{x}(0,t) =\bm{r}(0)+ \sin\left(\frac{t}{R}\right)\,\hat{\bm{e}}_2+R\left[\frac{2R}{a}\sin\left(\frac{a}{2R}\right)-\cos\left(\frac{t}{R}\right)\right]\hat{\bm{e}}_3 \quad \text{and}\quad \bm{x}_s
    \cdot\hat{\bm{e}}_3=0, \subtag{a,b}
\end{equation}
\end{subequations}
where $a=p(0)$. For this to be consistent with the small strain assumption, we need $a/(2R)\ll 1$. Comparing this with \eqref{eq:positionasym} and expanding in $a/(2R)$  yields the leading-order boundary conditions
\begin{subequations}\label{eq:clampingbcs}
\begin{gather}
u(0, t)=w(0,t)\phi(0), \quad v(0, t)=-\frac{t^3}{3! R^2}, \quad w(0,t)=\frac{1}{R}\left(\frac{t^2}{2}-\frac{a^2}{4!}\right),\quad \phi(0)=- \frac{\partial w}{\partial s}(0,t).\subtag{a--d}
\end{gather}
\end{subequations}
With this, we now turn to minimizing the energy for the intrinsic equations.

\paragraph{Center of mass strain equation}  Since   $\varepsilon(s)$ only appears in the longitudinal strain, $E_{ss}$, it immediately follows from varying  the energy, \eqref{eq:asymenergy}, with respect to  $\varepsilon$ that 
\begin{equation}\label{eq:tssavg}
    \overline{T_{ss}}=\overline{Y(E_{ss}+\nu E_{tt})}=0,
\end{equation}
at equilibrium --- i.e., the cross-sectional average of the longitudinal stress is zero. Inserting the strains from \eqref{eq:stretchstrains} provides an equation for $\varepsilon(s)$.

\paragraph{Center of mass slope equation}  Similarly, since $\phi(s)$ only appears in the $E_{ss}$ and $B_{ss}$ strains (as well as the gravitational potential energy), variations in $\phi$  yield the Euler--Lagrange equation
\begin{equation}\label{eq:phODE}
   \frac{\dd}{\dd s}\left[p(s)\Big(\overline{M_{ss}-w(s,t)T_{ss}}\Big)\right]=\rho g h I(s)  \cos\phi(s),
\end{equation}
for the integrated profile $I(s)=\int_s^Lp(\xi)\,\dd \xi$,  $M_{ss}=D(B_{ss}+\nu B_{tt})$, and $T_{ss}=Y(E_{ss}+\nu E_{tt})$. This provides a second-order ODE for $\phi(s)$, which is subject to
\begin{subequations}\label{eq:phBCs}
\begin{equation}
   \phi+\overline{\partial_sw}=0\enspace \text{at}\enspace s=0 \quad\text{and}\quad \overline{M_{ss}-w(s,t)T_{ss}}=0 \enspace \text{at}\enspace s=L. \subtag{a,b}
\end{equation}
\end{subequations}
Here, the condition at $s=L$ is the natural boundary condition that follows from the variation in $\phi$, while the condition at $s=0$  imposes  the plate is  horizontally clamped, i.e.,  \myeqref{eq:positionasym}{b}. Note that if $w=0$ (i.e.~the  vertical deflection is given purely by the center of mass), then $M_{ss}=D \phi'(s)$ and the BVP, \eqref{eq:phODE} subject to \eqref{eq:phBCs}, becomes
\begin{equation}\label{eq:phODE2}
    D\left[p(s)\phi'(s)\right]'=\rho g h I(s) \cos\phi(s)\quad \text{subject to} \quad \phi(0)=\phi'(L)=0,
\end{equation}
which is the equation for a heavy elastica with a non-uniform profile \cite{Wang86}.

Variations in the deformations $u(s,t)$, $v(s,t)$, and  $w(s,t)$ will lead to PDEs. Before considering these, we shall reduce the system further by posing an ansatz for the cross-section's profile.

\subsection{Shallow cylinder approximation}
For small clamping curvatures, $a/(2R)\ll1$, we expect the out-of-plane profile of the plate to not deviate far from the shallow cylinder approximation at the clamping edge, i.e., \myeqref{eq:clampingbcs}{c}. Comparisons to the finite element simulations confirm that this is true away from the possible buckling region (see \S\ref{AnsatzCheck}). We shall, thus, pose an ansatz for the out-of-plane deformation of the form
\begin{equation}\label{eq:wansatz}
    w(s,t)=c(s)\left(\frac{t^2}{2}-\frac{p(s)^2}{4!}\right),
\end{equation}
where $c(s)$ is the transverse curvature, and the constant term in $t$ is chosen so that $\overline{w(s,t)}=0$. With this ansatz, the clamping conditions \myeqref{eq:clampingbcs}{c,d} become
\begin{subequations}\label{eq:cclamp}
\begin{equation}
    c(0)=1/R,\quad c'(0)=0,\quad \Phi(0)=0,\subtag{a--c}
\end{equation}
\end{subequations}
where $\Phi(s)\coloneqq \phi(s)-\left[c(s)p(s)^2\right]'/4!$ is the slope along $t=0$.

The in-plane displacements, $u$ and $v$, then come from assuming that the transverse strains, $E_{st}$ and $E_{tt}$, identically vanish. Although this choice is not consistent with the stress-free boundary conditions that are typically imposed on the free lateral edges of the plate, $t=\pm p(s)/2$, we find that these boundary conditions have a negligible effect on the deflection of the plate due to Saint-Venant's Principle. Furthermore, we find that other simplifications  (for example, setting $T_{tt}=T_{st}=0$), lead to qualitatively similar results, but yield more complex equations. Inserting \eqref{eq:wansatz} into \myeqref{eq:stretchstrains}{b,c} with $E_{tt}=E_{st}=0$ and solving for $u$ and $v$ yields the ansatzes
\begin{subequations}
\begin{equation}
u(s,t)=\frac{c(s)[c(s)p(s)^2]'}{4!} \left(\frac{t^2}{2}-\frac{p(s)^2}{4!}\right)-c(s)c'(s)\left(\frac{t^4}{4!}-\frac{p(s)^4}{16\times 5!}\right)\quad\text{and}\quad v(s,t) = -c(s)^2 \frac{t^3}{3!}.
\end{equation}
\end{subequations}
Note that these are consistent with the clamping conditions \myeqref{eq:clampingbcs}{a,b}, provided \eqref{eq:cclamp} hold. 

With this ansatz, the non-zero stretching strain is
\begin{equation}\label{eq:strainansatz}
E_{ss}=-c(s)\Phi'(s) \left(\frac{t^2}{2}-\frac{p(s)^2}{4!}\right)+\left[2c'(s)^2-c(s)c''(s)\right]\left(\frac{t^4}{4!}-\frac{p(s)^4}{16\times 5!}\right), 
\end{equation}
where we have already chosen $\varepsilon$ so that $\overline{ E_{ss}}=0$, i.e., \eqref{eq:tssavg},
while the bending strains are
\begin{subequations}\label{eq:bendstrainansatz}
\begin{gather}
    B_{ss} =\Phi'(s)+ c''(s)\frac{t^2}{2},\quad
    B_{tt} = c(s),\quad   B_{st} =B_{ts} =c'(s)t.\subtag{a--c}\end{gather}
\end{subequations}

Inserting \eqref{eq:strainansatz} and  \eqref{eq:bendstrainansatz} into  \eqref{eq:asymenergy}, yields an energy dependent on the unknown curvature $c(s)$ and slope $\phi(s)$. Variations in $\phi$ yields the ODE \eqref{eq:phODE}, variations in $c(s)$, yield the Euler-Lagrange equation
\begin{equation}
    \frac{\dd^2}{\dd s^2}\left[p(s)\Bigg(\overline{\frac{\partial B_{ij}}{\partial c''}M_{ij}+\frac{\partial E_{ij}}{\partial c''}T_{ij}}\Bigg)\right]-\frac{\dd}{\dd s}\left[p(s)\Bigg(\overline{\frac{\partial B_{ij}}{\partial c'}M_{ij}+\frac{\partial E_{ij}}{\partial c'}T_{ij}}\Bigg)\right]
    +p(s)\Bigg(\overline{\frac{\partial B_{ij}}{\partial c}M_{ij}+\frac{\partial E_{ij}}{\partial c}T_{ij}}\Bigg)=0.
\end{equation}
This is subject to the clamping conditions \eqref{eq:cclamp} at $s=0$ and the natural boundary conditions
\begin{subequations}
\begin{gather}
    \overline{\frac{\partial B_{ij}}{\partial c''}M_{ij}+\frac{\partial E_{ij}}{\partial c''}T_{ij}}=0,\quad\text{and} \quad
    \frac{\dd}{\dd s}\Bigg(\overline{\frac{\partial B_{ij}}{\partial c''}M_{ij}+\frac{\partial E_{ij}}{\partial c''}T_{ij}}\Bigg)=\overline{\frac{\partial B_{ij}}{\partial c'}M_{ij}+\frac{\partial E_{ij}}{\partial c'}T_{ij}},\subtag{a,b}
\end{gather}
\end{subequations}
at $s=L$.

If we introduce superscripts to denote  a function's moments,
\begin{subequations}
\begin{gather}
    f^0(s)\coloneqq \int_{p(s)/2}^{p(s)/2}f(s,t)\,\dd t,\quad f^1(s)\coloneqq \int_{p(s)/2}^{p(s)/2} t f(s,t)\,\dd t,\subtag{a,b}\\ f^2(s)\coloneqq \int_{p(s)/2}^{p(s)/2}\left(\frac{t^2}{2}-\frac{p(s)^2}{4!}\right)f(s,t)\,\dd t,\quad f^4(s)\coloneqq \int_{p(s)/2}^{p(s)/2}\left(\frac{t^4}{4!}-\frac{p(s)^4}{16\times 5!}\right)f(s,t)\,\dd t,\subtag{c,d}
\end{gather}
\end{subequations}
then the  ODEs for  $c(s)$ and $\phi(s)$  can be written as 
\begin{subequations}
\begin{gather}
\frac{\dd}{\dd s}\left[M^0_{ss}(s)-c(s)T^2_{ss}(s)\right]''=\rho g h I(s)  \cos\phi(s),\\
\begin{split}
    \text{and}\quad \frac{\dd^2}{\dd s^2}\left[M_{ss}^2(s)+\frac{p(s)^2}{4!}M_{ss}^0(s)-c(s)T_{ss}^4(s)\right]-2\frac{\dd}{\dd s}\left[M_{st}^1(s)+2c'(s)T_{ss}^4(s)\right]\\
    +M_{tt}^0(s)-c''(s)T_{ss}^4(s)-\Phi'(s)T_{ss}^2(s)=\frac{p(s)^2}{4!}\frac{\dd^2}{\dd s^2}\left[M^0_{ss}(s)-c(s)T^2_{ss}(s)\right]'',
    \end{split}
\end{gather}
\end{subequations}
for $0\leq s\leq L$, which are subject to \eqref{eq:cclamp} on $s=0$  and the natural boundary conditions on $s=L$:
\begin{subequations}
\begin{gather}
M_{ss}^0(s)=c(s)T_{ss}^2(s),\\
    M_{ss}^2(s)+c(s)\frac{p(s)^2}{4!}T_{ss}^2(s)-c(s)T_{ss}^4(s)=0,\\
     \frac{\dd}{\dd s}\left[ M_{ss}^2(s)+c(s)\frac{p(s)^2}{4!}T_{ss}^2(s)-c(s)T_{ss}^4(s)\right]=2M_{st}^1(s)+4c'(s)T_{ss}^4(s).
\end{gather}
\end{subequations}
 Here, the non-zero  moments of the stresses are  
\begin{subequations}
\begin{align}
    M_{st}^1(s)&=2D(1-\nu)\frac{c'(s)p(s)^3}{4!},\\
    M_{tt}^0(s)&=D\left[c(s)p(s)+\nu \Phi'(s)p(s)+\nu c''(s)\frac{p(s)^3}{4!}\right],\\ 
    M_{ss}^0(s)&=D\left[\Phi'(s)p(s)+c''(s)\frac{p(s)^3}{4!}+\nu c(s)p(s)\right],\\
   M_{ss}^2(s)&=D\frac{p(s)^5}{6!}c''(s),\\
    T_{ss}^2(s)&=Y\left[-c(s)\Phi'(s)\frac{p(s)^5}{6!}+[2c'(s)^2-c(s)c''(s)]\frac{p(s)^7}{8!}\right],\\
    T_{ss}^4(s)&=Y\left[-c(s)\Phi'(s)\frac{p(s)^7}{8!}+[2c'(s)^2-c(s)c''(s)]\frac{7  p(s)^9}{4 \times10!}\right].
\end{align}
\end{subequations}

The resulting BVP was solved using the BVP solver \texttt{bvp4c} in \textsc{Matlab}. The plate drop, $d$, is then computed using \eqref{eq:positionasym}, i.e.
\begin{equation}
    d = \hat{\bm{e}}_3\cdot\bm{x}(L,0)- \hat{\bm{e}}_3\cdot\bm{x}(0,0) = \int_0^L\sin\phi(s)\,\dd s+w(L,0)\cos\phi(L)-w(0,0)\cos\phi(0)+\Oh(H^2).
\end{equation}

\vspace{-.5cm} 
\section{Motivation/validation of reduced-order ansatzes}\label{AnsatzCheck}
\begin{figure}[ht]
\centering
\includegraphics[width=.95\textwidth]{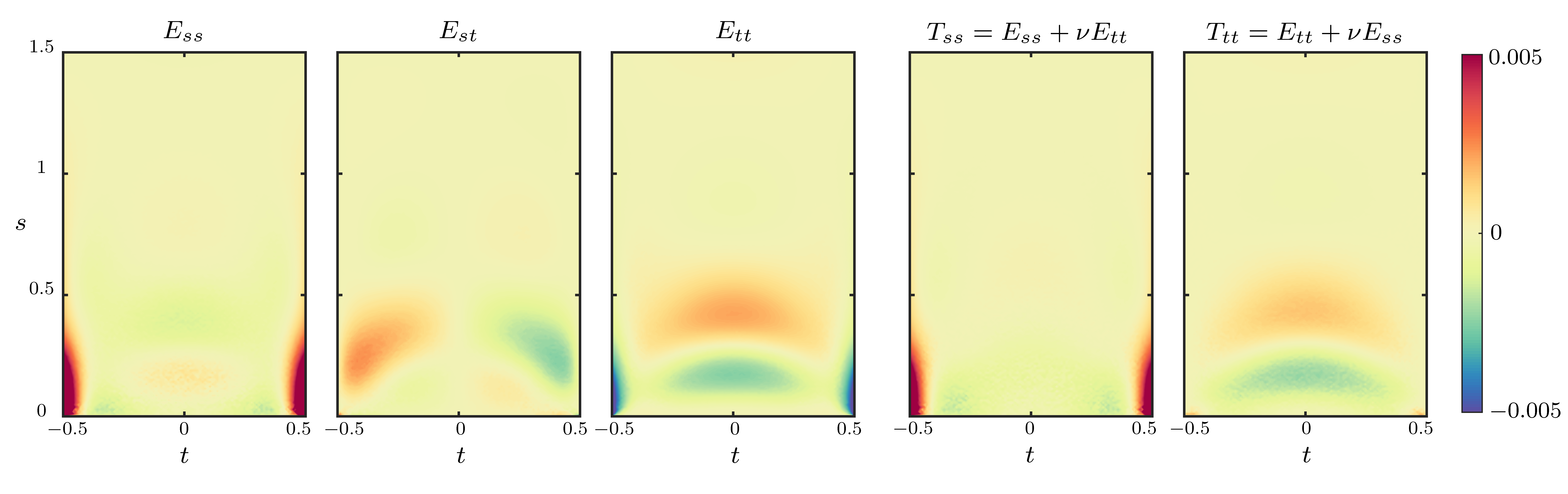}
\caption{In-plane strains, $E_{ss}$, $E_{st}$, and $E_{tt}$, and stresses, $T_{ss}$ and $T_{tt}$, for a rectangular sheet of length $L/a = 1.5$, clamping curvature $a/R = 0.5$, and thickness $h/a = 0.03175$, without gravity, computed using finite element simulations.  Save for a small boundary layer near the clamped edge, strains and stresses are all small. 
}
\label{FigureS3}
\end{figure}

\begin{figure}[ht]
\centering
\includegraphics[width=.9\textwidth]{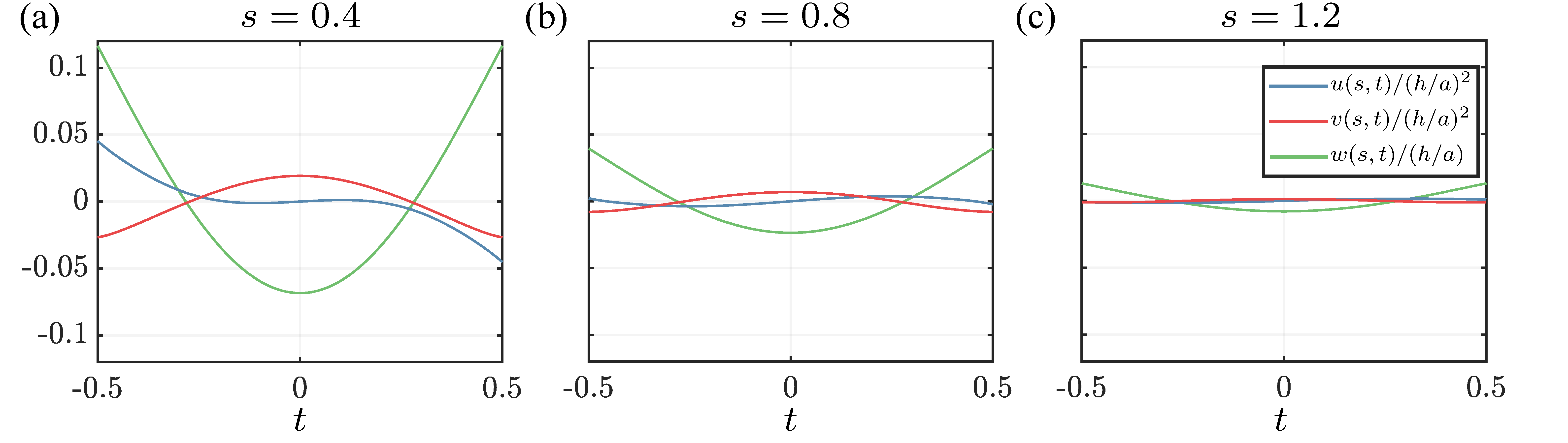}
\caption{Deformations $u(s,t)/(h/a)^2$, $v(s,t)/(h/a)^2$, and $w(s,t)/(h/a)$ of the mid-surface at three different cross sections ($s/a\in\{0.4, 0.8, 1.2\}$) of a rectangular sheet of scaled length $L/a = 2$, clamping curvature $a/R = 0.5$, and thickness $h/a = 0.03175$, without gravity. The out-of-plane deformation, $w$, scales with $h/a$, and is quadratic in $t$ at leading order. The in-plane deformations, $u$ and $v$, enter at $\Oh(h/a)^2$.}
\label{FigureS4}
\end{figure}

\vspace{-.1cm} 
Motivation of the ansatzes used in the reduced-order model are provided in Figs.~\ref{FigureS3} and \ref{FigureS4}. Figure~\ref{FigureS3} shows the in-plane strains $E_{ss}$, $E_{st}$, $E_{tt}$, and stress $T_{ss}$, $T_{tt}$ for a rectangular sheet of length $L/a = 1.5$, clamping curvature $a/R = 0.5$, and thickness $h/a = 0.03175$ without gravity, computed using finite element simulations. Save for a small boundary layer near the clamped edge, strains and stresses are all small.

Finally, Fig.~\ref{FigureS4} shows the scaled deformations $u(s,t)/(h/a)^2$, $v(s,t)/(h/a)^2$, and $w(s,t)/(h/a)$ of the mid-surface at three different cross sections, $s/a\in\{0.4, 0.8, 1.2\}$, of a rectangular sheet with scaled length $L/a = 2$, clamping curvature $a/R = 0.5$, and thickness $h/a = 0.03175$, without gravity. The out-of-plane deformation, $w$, scales with $h/a$, and is quadratic in $t$ at leading order. The in-plane deformations, $u$ and $v$, are thus seen to enter at $\Oh(h/a)^2$.

\end{document}